\documentclass[aps,prb,floatfix,amsmath,amssymb,amsfonts,showpacs,groupedaddress,twocolumn,10pt]{revtex4-1}

\usepackage{keyval} \usepackage{dcolumn}
\usepackage[]{amsmath}
\usepackage{xspace}

\usepackage{graphicx}
\usepackage{natbib}

\usepackage{bm}

\usepackage[usenames,dvipsnames]{color}

\usepackage[normalem]{ulem}
\setkeys{Gin}{width=0.90\columnwidth}
\graphicspath{{figs/}{:Figs:}} 

\newcommand{\muB}{\ensuremath{\mu_{\mathrm{B}}}}

\newcommand{\SRO}{SrRuO$_{3}$\xspace}
\newcommand{\V}{$V^{\text{calc}}_{\text{eq}}$\xspace }

\begin{document}

% Title of paper
%\title{\emph{Working title: }Assesment of the magnetic properties and electronic structure of \SRO using LDA, LDA+$U$ and LDA+DMFT}
%\title{ DMFT treatment of \SRO}
%\title{Electronic structure of \SRO~from dynamical mean field theory}
\title{Electronic structure, cohesive properties and magnetism of \SRO}
\author{Oscar Gr\aa n\"as}
\affiliation{Department of Physics and Astronomy, University
  of Uppsala}

\author{Igor Di Marco}
\affiliation{Department of Physics and Astronomy, University
  of Uppsala}

\author{Olle Eriksson}
\affiliation{Department of Physics and Astronomy, University
  of Uppsala}

\author{Lars Nordstr\"om}
\affiliation{Department of Physics and Astronomy, University
  of Uppsala}
  
\author{Corina Etz}
\affiliation{Department of Physics and Astronomy, University
  of Uppsala}

\date{\today}

\begin{abstract}
We have performed an extensive test of the ability of density functional theory within several approximations for the exchange-correlation functional, local density approximation+Hubbard $U$ and local density approximation + dynamic mean field theory to describe magnetic and electronic properties of \SRO. We focus on the ferromagnetic phase, illustrating differences between the orthorhombic low temperature structure vs the cubic high temperature structure. We assess how magnetism, spectral function, and cohesive properties are affected by methodology, on-site Hubbard $U$ and double counting corrections. Further, we compare the impact of the impurity solver on the quasiparticle weight $Z$, which is in turn compared to experimental results. The spectral functions resulting from the different treatments are also compared to experimental data. Finally, the impact of spin-orbit coupling is studied, allowing us to determine the orbital moments. In the orthorhombic phase the orbital moments are found to be tilted with respect to the spin moments, emphasising the importance of taking into account the distortion of the oxygen octahedra. 
\end{abstract}

\pacs{75.47.Lx,71.15.Mb,71.27.+a}

\maketitle

\section{Introduction}
%give the interesting story of the paper in introduction !!!! \\
The ruthenates have been the object of various experimental and theoretical investigations. This interest is stimulated by the fact that the layered ruthenates exhibit an impressive list of exotic phenomena. For example Sr$_{3}$Ru$_{2}$O$_{7}$ where meta-magnetism is known to occur,~\cite{Lee:2009hm}  Sr$_{2}$RuO$_{4}$ with unusual $p$-wave superconductivity~\cite{Maeno1994,Mravlje:2011gk} or long-range ferromagnetic order (rare in 4$d$-based materials) in \SRO. The latter is a versatile compound whose properties can be drastically changed by doping or strain. For example Ca doping on the Sr sites leads to a poor metal, possibly non-Fermi-liquid behaviour, with no trace of magnetism,~\cite{Klein:1999tz} while doping Cu on the Ru sites creates an insulating spin glass.~\cite{Mangalam:2009dc} %In the latter the magnetic properties change drastically by small changes in the electronic structures.  
The double-perovskite Sr$_{2}$RuYO$_{6}$ has basically the same crystal structure as \SRO  but one obtains an antiferromagnetic configuration by substituting every second Ru in Sr$_{2}$RuYO$_{6}$ with Y (Ref.~\onlinecite{Mazin:1997es}). Hence the ferromagnetic state is unstable under small changes in the electronic structure, but still robust in the temperature range below $\approx 160 \text{ K}$. Also, intrinsic defects such as Ru vacancies are shown to have a big impact on magnetic properties.~\cite{Dabrowski:2004ix} \SRO also shows an unusually  high magnetic anisotropy energy for an almost cubic material.~\cite{Kanbayasi:1976kn,Mazin:1997es} Besides varied and interesting properties, the ruthenates are also good candidates for oxide electronics,~\cite{Koster:2012gr} either alone or in combination with other complex oxides (e.g. in superlattices~\cite{Ziese:2010fk}). Even more, the hypothesis that \SRO might support the existence of magnetic monopoles in $k$-space~\cite{Fang:2003uq} has been formulated.

Experimental measurements of the magnetic moment vary from $0.86~\muB$ to $1.6~\muB$, from e.g. Refs.~\onlinecite{Felner:2006hm,Kanbayasi:1976kn}, showing the delicate nature of the magnetic state. This large variation has been attributed to the difficulty of saturating the moment, due in part to the high magnetic anisotropy and in part to the challenging task of synthesising single domain structures.~\cite{Rondinelli:2008jy} One cannot neglect the dependence of the magnetic properties on the sample quality. %In conclusion, even the sample quality is a non-negligible 
%This large variation has been attributed to the anomalously high magnetic anisotropy leading to difficulties to saturate the moment due to the difficulty to synthesize single domain structures~\cite{Rondinelli:2008jy}, so we cannot neglect either the dependence of the properties on the sample quality. 

Examining results from electronic structure calculations one also sees a variety of moments ranging from 0.5 to 2.0 $\muB$/Ru atom.~\cite{Jakobi:2011cs,Jeng:2006iq} In order to give a correct description of the properties of \SRO, several computational tools have been applied. These include LDA/GGA,~\cite{Mazin:1997es,Rondinelli:2008jy} LDA+$U$,~\cite{Jeng:2006iq} pseudo-SIC,~\cite{Rondinelli:2008jy} GWA + LDA+$U$,~\cite{Hadipour:2011ew} and recently LDA+DMFT.~\cite{Jakobi:2011cs} 
One may observe from previous studies that the DFT description of \SRO strongly varies with respect to the (i) type of exchange-correlation (E$_{xc}$) functional used or (ii) the way these functionals are implemented. 
The main controversy about \SRO is the role of strong correlation in various properties. Infrared conductivity hints of non-Fermi liquid properties~\cite{Kostic1998} and enhanced effective mass is detected in various experiments.~\cite{Cox:1983wx,Cao:1997id,Ahn:1999ks} Yet many properties seem to be well described within Kohn-Sham density functional theory, although very sensitive to the choice of exchange-correlation functional used.
%\comment{Removed the part about single slater determinants, not quite true for the electron system, only for the quasiparticles as ref. 1 points out.}
The DFT+SIC (self interaction correction) treatment of \SRO has proven to introduce an unphysical localisation of the Ru 4$d$ states, leading to half-metallicity and an overestimation of magnetic moments, approaching the results given by DFT+$U$, in the limit of large Hubbard $U$. With a more sophisticated method, such as LDA+ dynamical mean field theory (DMFT), it becomes interesting to investigate in detail the electronic structure and the effect of on-site correlations in \SRO. Indeed, DMFT studies of models mimicking the cubic phase of \SRO highlight the importance of on-site correlation effects, however not driven by the proximity to the Mott insulator, but instead by the Hund's coupling.~\cite{deMedici:2011uq,Georges:2013}

% Our contribution to the scientific community interested in the correct description of the physics of strontium ruthenates is a ...complex/complete/exhaustive..... study based on dynamical mean-field theory.
%What we bring new is a 

 The present study is focused on the investigation of the electronic structure and magnetic properties of \SRO, both in the cubic and orthorhombic phase, by means of first principles DFT-based methods, in a full-potential geometry with emphasis on a many-body approach to correlated  systems, i.e. the DMFT.~\cite{Vollhardt:2012fk,Kotliar2006} In contrast to previous LDA+DMFT studies~\cite{Jakobi:2011cs} we employ a full potential electronic structure method~\cite{wills:2010th} in which the first implementation of LDA+DMFT was reported in Ref.~\onlinecite{Grechnev:2007en}, while details on the full charge self consistency were reported in Ref.~\onlinecite{Granas:2012hg}. 
 The purpose of this study is to evaluate what method is most suitable for working with \SRO in its low temperature phase. In order to achieve our aim, we compare the calculated quantities with experimentally measured data, such as magnetic moments, effective masses and photoemission spectra assessing the sensitivity to exchange correlation within DFT, followed by accurate full-charge self-consistent LDA+DMFT calculations, where $U$ is varied within a reasonable range. In this way we make a critical assessment of the method that describes most accurately the properties of the \SRO.

% We asses the sensitivity of the magnetic moment with respect to exchange and correlation within DFT,  

%The rigorous treatment of on the low-temperature phase,

% method  which is important due to the sensitivity of the system to changes in crystal and/or electronic structure.

\section{Methodology}
The DFT implementation used here employs a full-potential geometry and a Linearized Muffin-Tin Orbitals (LMTO) basis set, as implemented in the RSPt code.~\cite{rspt,Wills:1987vt,wills:2010th} For the DFT part of the calculations we treat the narrow semi-core bands with a double $\kappa$-basis set. The $s$-, $p$- and $d$-valence bands are treated with a triple $\kappa$-basis. The LDA+DMFT calculations are run with a double $\kappa$-basis in order to reduce the matrix dimensions and speed up the calculation. Tests with double and triple $\kappa$-basis were performed for all volumes in order to assess that the quality of the DMFT results is not affected by the reduced basis. We perform calculations for \SRO in both the cubic and orthorhombic phases. The Brillouin zone sampling for the orthorhombic phase is converged with respect to the equilibrium volume, magnetic moment, bulk modulus, spectral function and quasiparticle weights  at $12\times12\times8$ $k$-points. The effects of spin-orbit interaction are more delicate and did not show convergence with a mesh of less than  $24\times24\times16$ $k$-points (in order to ensure convergence, meshes up to $30\times30\times24$ $k$-points were considered). The cubic phase is converged at a mesh of $24\times24\times24$ $k$-points. For the DFT calculations we  used the tetrahedron method with Bl\"ochl's corrections~\cite{Blochl:1994vg} for integrations. As a starting point for the LDA+DMFT calculation we use the parametrization by von Barth and Hedin (vBH72). The final result will of course be somewhat dependent on the starting point. The LDA+DMFT is implemented in the Matsubara formalism,~\cite{Grechnev:2007en,Granas:2012hg} hence Fermi smearing is implicit. An electronic temperature of $110 \text{ K}$ is used. We test five commonly used exchange-correlation functionals (E$_{xc}$), the LDA parametrizations by von Barth and Hedin (vBH72),~\cite{Barth:1972un} Perdew and Wang (PW91),~\cite{Perdew:1992vq} and Perdew and Zunger (PZ81)~\cite{Perdew:1981wm} as well as the GGA parametrizations by Armiento and Mattsson (AM05)~\cite{Armiento:2005fk} and Perdew, Burke and Enzerhof (PBE96).~\cite{Perdew:1996uq} 

% The resulting moments for the experimental structure are shown in Table~\ref{tab:LDAmoments}.

The impurity problem in the LDA+DMFT scheme is solved using the spin-polarized T-matrix fluctuation exchange solver (SPTF).~\cite{Pourovskii:2005uw} Two versions are tested, one where the diagrams are calculated with the unperturbed Green's function (G$_{0}$) and one where the perturbation expansion is performed with a partially renormalized Green's function (G$_{\mathrm{HF}}$) which consistently shows improvement in the description of magnetic properties.~\cite{newFLEX} The SPTF solver, based on the fluctuation exchange (FLEX) approximation,~\cite{Bickers:1989fk} works in the weak to intermediate coupling regime, when the ratio between the average local Coulomb interaction $U$ and the average bandwidth $W$ is smaller than one.~\footnote{Notice that the significance of the $U/W$ ratio to classify the strength of correlations is somewhat diminished for multi-orbital systems where the corresponding orbitals are associated to different bandwidths and different matrix elements of the Coulomb interaction. In the main text we refer to this ratio only to to give an idea of the regime of applicability of our solver in terms of standard classifications.} In \SRO the total 4d-bandwidth is over $\approx$ 12 eV, with a distinct feature of a significantly narrower t$_{2g}$-band pinned at the Fermi level. The $U$ values considered in this study do not overcome 4~eV, and therefore we are in the regime where the SPTF solver is valid. As double counting we remove the static part of the self-energy.~\cite{Lichtenstein:2001uq} We also calculate the spectral properties and magnetic moments using LDA+$U$, testing both the fully localized limit (FLL) and the around mean field (AMF) double countings. The 4-index rotationally invariant matrix of the Coulomb interaction for the Ru-4d electrons is constructed from two parameters $U$ and $J$, as described elsewhere.~\cite{Grechnev:2007en,Granas:2012hg} The value of $J$ is set to 0.6~eV according to constrained random-phase approximation calculations on Ru metal by Sasioglu {\itshape et al.}~\cite{Sasoglu:2011ch} Unfortunately, the same procedure cannot be used for $U$, which is more sensitive to the details of the electronic environment (different screening of valence electrons), and also strongly dependent on the basis set used for the local orbitals. In this work, we perform all simulations for $U=2$, 3 and 4 eV. These values are chosen to be compatible with the value $U = 4 \text{ eV}$ obtained for Ru metal~\cite{Sasoglu:2011ch} and by considering that SPTF tends to overestimate the strength of correlations with respect to numerically exact solvers, since inter-orbital (Ru-4d) screening is only partially considered in perturbative approaches. Although all properties presented in this study have been calculated for all values of $U=2$, 3 and 4 eV, only selected data are reported here, for sake of readability. In particular the volume trends, magnetic moments and effects of spin-orbit coupling are reported for $U= 3 \text{ eV}$, which seems to offer the best comparison with experimental data. Changes of ground-state and spectral properties with respect to $U$ are investigated in separate sections. Concerning the spectral properties, curves for $U= 3 \text{ eV}$ are omitted from the plots, for sake of clarity, as explained in the corresponding section.

\section{Results}
The phase diagram of \SRO shows that for temperatures up to $825 \text{ K}$ the compound is present in the orthorhombic structure, with a ferromagnetic order that prevails up to $T_{\text{c}}=160 \text{ K}$. In the temperature range between $825 \text{ K}$ and $945 \text{ K}$, the crystal is tetragonal and becomes cubic for temperatures above $945 \text{ K}$. %, the structure of \SRO is cubic. 
In the following, we focus on the orthorhombic phase, while making comparisons to the cubic structure with ferromagnetic arrangement of magnetic moments. In this way, we wish to determine and highlight the effect of the octahedral distortions and tilting on the properties of \SRO. The experiments show no sign of ferromagnetism present in the cubic phase, although they indicate that~\cite{Shai2013} local moments persist above $T_{\text{c}}$. These moments are most likely to contribute significantly to the physical properties of the cubic phase. %\textit{The theoretical work by Rondinelli {\itshape et al.}~\cite{Rondinelli:2008jy} shows that the ferromagnetic configuration is lower in energy.}To properly describe the cubic phase one should therefore resort to DFT+DMFT or at least disordered local moment treatment.   %% (Ref.~\onlinecite{Shai2013}) 

The orthorhombic cell is described according to recent neutron diffraction experiments by Bushmeleva {\itshape et al.}~\cite{Bushmeleva:2006jn} Their experiments show almost equal bond distances of $\approx 1.986 \text{ \AA}$ for all independent Ru-O bonds. This indicates the absence of Jahn-Teller distortions in the RuO$_{6}$ octahedra, confirming a low-spin state of the Ru$^{4+}$ ion. The two independent Ru-O-Ru angles remain close to 162$^{\circ}$ for the whole temperature range between $1.5 \text{ K}$ and $290\text{ K}$. The difficulty to achieve a single magnetic domain is manifested in a severe overlap between diffraction peaks and in consequence the direction of the magnetic moment in the cell cannot be determined unambiguously.

%As mentioned before, \SRO has been studied theoretically by means of a multitude of methods. The main reason is that in spite of the 4$d$-magnetism, the compound has been %considered to exhibit strong electron-electron correlations. Recently it has been suggested (both by experiment~\cite{Maiti2005,Lee:2001kx} and theory~\cite{Etz:2012dj}) %that the electronic correlations in \SRO are weak. In the following we wish to provide an extensive analysis of the properties of this compound and, by comparing the calculated %results to experimental data, make an assessment on the degree of electron correlations and the best suited method to treat the electronic structure of \SRO and related %ruthenates. 

As mentioned before, \SRO has been studied theoretically by means of a multitude of methods. Clean \SRO samples are known to exhibit Fermi-liquid  behavior at low temperatures. Although the possibility of a bad metal phase (Hund's metal) at higher temperature is argued from model calculations on a $t_{2g}$ manifold without long range magnetic order ~\cite{deMedici:2011uq,Georges:2013}, recently it has been suggested (both by experiment~\cite{Maiti2005,Lee:2001kx} and theory~\cite{Etz:2012uq}) that electronic correlations in \SRO are sufficiently weak for conventional DFT functionals to reproduce the ground state. In the following we wish to provide an extensive analysis of the properties of this compound and, by comparing the calculated results to experimental data, make an assessment on the degree of electron correlations and the best suited method to treat the electronic structure of \SRO, as well as determining to what extent a dynamic treatment of the exchange-correlation problem influences its properties.

\subsection{Volume trends}
Due to the complexity of the \SRO phase diagram, it is far from sufficient to compare calculated values of magnetic moments with the measured values, without making a careful analysis regarding the volume dependence of the moments. In the following we will show that there is quite a strong volume dependence of the magnetic moments.  This might in fact be one of the reasons for the wide range of calculated magnetic moments, available in the literature.

\begin{figure}[htb]
\begin{center}
%\hspace{-1cm}
\begin{tabular}{c}
\includegraphics[width=0.5\textwidth,clip]{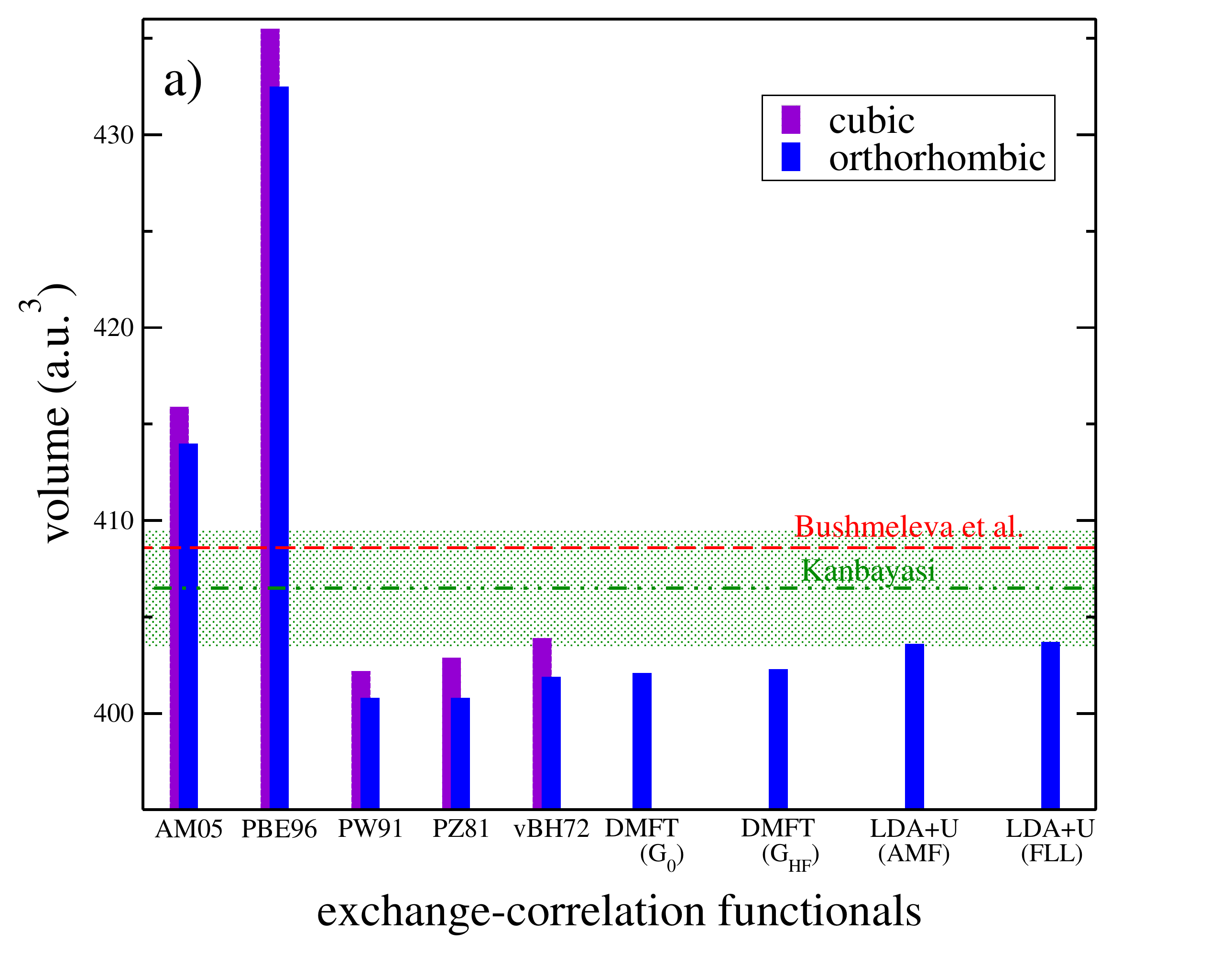} \\ 
\includegraphics[width=0.5\textwidth,clip]{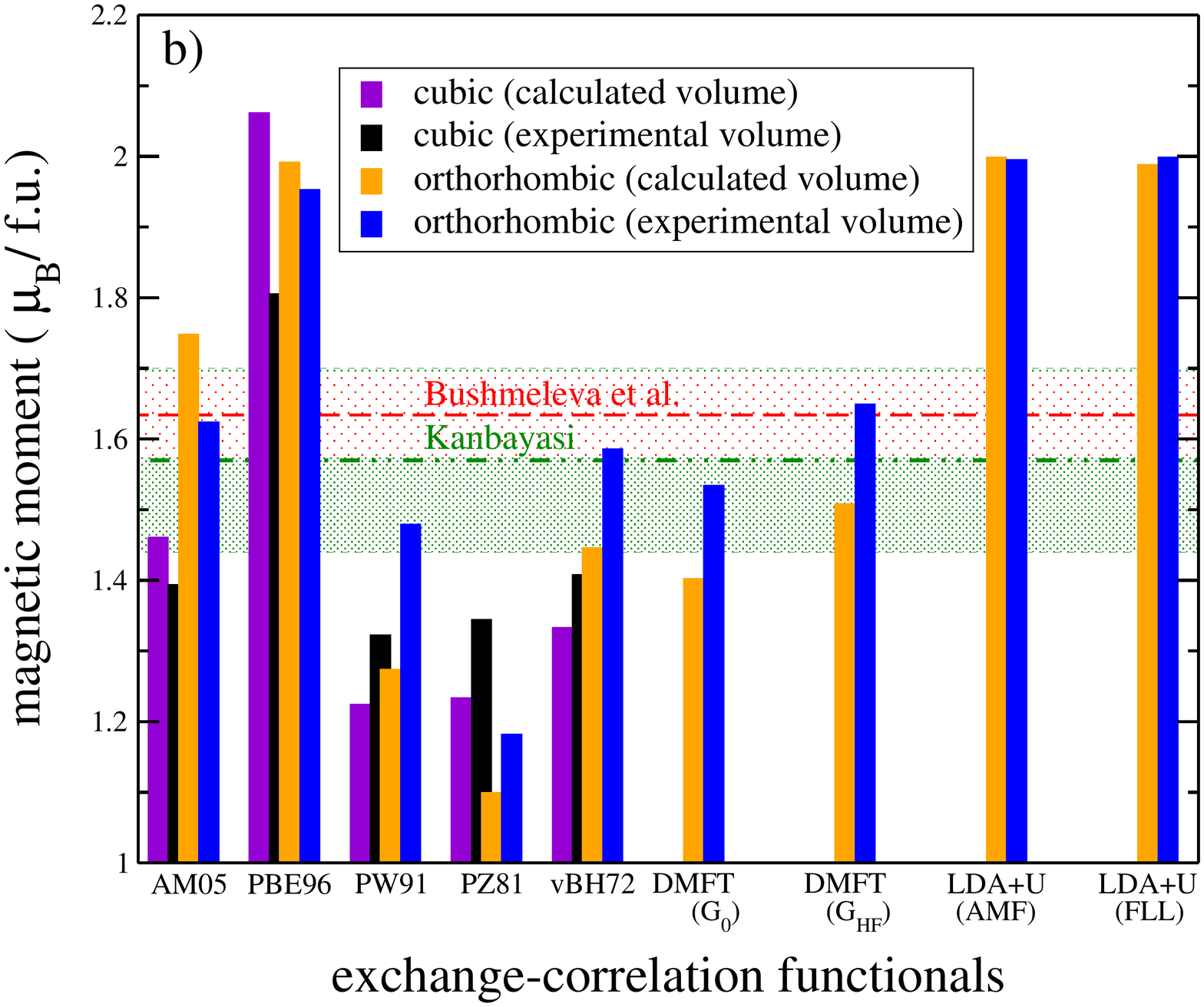} \\
\end{tabular}
\caption{(colour online) (a) Calculated equilibrium volumes (\V) determined by different methods and functionals (LDA, GGA, LDA+DMFT and LDA+$U$ with $U=3 \text{ eV}$). % Energy vs volume for a) the orthorhombic structure and b) the cubic structure for different LDA/GGA functionals, LDA+DMFT and LDA+U, with $U$=3.0 eV. 
(b) Spin magnetic moments per formula unit calculated for both the experimental volume~\cite{Bushmeleva:2006jn} as well as for the calculated equilibrium volumes. Both cubic (brown and black bars) and orthorhombic (blue and orange bars) structures have been considered. The experimentally determined values for saturation volume for single crystal bulk samples (Fig. 1a) and their corresponding magnetic moments (Fig. 1b) are marked by red dashed lines~\cite{Bushmeleva:2006jn} and green dash-dotted lines.~\cite{Kanbayasi:1976kn} The red (green) shaded area represents the interval associated to the experimental errors for Ref.~\onlinecite{Bushmeleva:2006jn} (Ref.~\onlinecite{Kanbayasi:1976kn}).} %For the LDA+$U$ and DMFT calculations a $U$= 3~eV was used.}
\label{fig:fig1}
\end{center}
\end{figure}

We start by determining the equilibrium volume for \SRO in both the orthorhombic as well as the cubic structure with ferromagnetic moment. The $a/b$ and $a/c$ ratios as well as the tilting and rotation angles for the orthorhombic phase are fixed to the experimental value. Fixing the structural parameters is expected to have a small impact on the equilibrium volume. It is known that the tilting and rotation angles do not change significantly for deviations within a few percent of the equilibrium volume~\cite{Zayak2006}, hence the fit is expected to be good at these volumes. The calculated equilibrium volumes (\V) determined within DFT and different approximations for the exchange-correlation as well as Hubbard-corrected LDA functionals and LDA+DMFT are summarised and compared to experimental data~\cite{Bushmeleva:2006jn,Kanbayasi:1976kn} in Fig.~\ref{fig:fig1} and Table~\ref{table1}. 
%\subsubsection{Experimental volume}
%

We have performed calculations for a range of volumes (for both orthorhombic and cubic phases), using the methodologies mentioned above. 
%The calculated values for the equilibrium volumes are summarised in Fig.~\ref{fig:fig1}. 
We fit the results to the universal equation of state by Vinet {\itshape et al.}~\cite{Vinet:1989fv} and from this we extract the equilibrium volume and bulk modulus (Table~\ref{tab:fig1}). 
The calculated volumes and bulk modulus are compared to experimental values for the orthorhombic phase. The experimental bulk modulus B$_0$ is $\approx 192(3) \text{ GPa}$, from Ref.~\onlinecite{Hamlin:2007ct}. The experimental volume estimated at $T = 290 \text{ K}$ is $\approx 408.582(7) \text{ a.u.}^{3}$ according to Bushmeleva {\itshape et al.}~\cite{Bushmeleva:2006jn} and $\approx 406.5 (18)\text{ a.u.}^{3}$ as estimated by Kanbayasi.~\cite{Kanbayasi:1976kn}

%modulus B$_0$ is $\approx 192 \pm 3 \text{ GPa}$, from Ref.~\onlinecite{Hamlin:2007ct}. The experimental volume estimated at $T = 290 \text{ K}$ is $\approx 408.582 \pm 0.007 \text{ a.u.}^{3}$ according to Bushmeleva {\itshape et al.}~\cite{Bushmeleva:2006jn} and $\approx 406.5 \pm 1.8 \text{ a.u.}^{3}$ as estimated by Kanbayasi.~\cite{Kanbayasi:1976kn}
%
%
%\hspace{-3cm}
\begin{table}[htdp]
\begin{center}
\begin{tabular}{|c|c||c|c|c|}
%\hline
%E$\mathrm{_{xc}}$ & $\langle S\rangle$/f.u Exp vol  & V$_{0}$ [a.u.$^{3}$]& $\langle S\rangle$/f.u & B0 [GPa]\\
\hline
\multicolumn{5}{|c|}{Orthorhombic structure} \\
\hline
E$\mathrm{_{xc}}$ & $\langle gS\rangle$/f.u. & V$_{0}$ [a.u.$^{3}$]& $\langle gS\rangle$/f.u & B$_0$ [GPa]\\
 & exp. vol. & & calc. vol. & \\
\hline

%\hline
%AM05       & 1.625 $\muB$& 413.957 & 1.749 $\muB$ & 178.2 \\
%PBE96      & 1.954 $\muB$& 432.465 & 1.993 $\muB$ & 161.9 \\
%PW91        & 1.481 $\muB$& 400.794 & 1.275 $\muB$ & 201.9 \\
%PZ81         & 1.183 $\muB$& 400.798 & 1.100 $\muB$ & 202.3 \\
%vBH72       & 1.587 $\muB$& 401.909 & 1.447  $\muB$ & 200.8 \\
%DMFT(G$_{0}$)  & 1.535 $\muB$& 402.124 & 1.403 $\muB$ & 188.8 \\
%DMFT(G$_{\text{HF}}$)  & 1.649 $\muB$& 402.317 & 1.509 $\muB$ & 185.0 \\
%LDA+$U$(AMF)   & 1.996 $\muB$& 403.555 & 1.989 $\muB$ & 201.2 \\
%LDA+$U$(FLL)   & 2.000 $\muB$& 403.669 & 2.000 $\muB$ & 202.6 \\
%\hline

AM05       & 1.6 $\muB$& 414.0 & 1.7 $\muB$ & 178.2 \\
PBE96      & 2.0 $\muB$& 432.5 & 2.0 $\muB$ & 161.9 \\
PW91        & 1.5 $\muB$& 400.8 & 1.3 $\muB$ & 201.9 \\
PZ81         & 1.2 $\muB$& 400.8 & 1.1 $\muB$ & 202.3 \\
vBH72       & 1.6 $\muB$& 401.9 & 1.4  $\muB$ & 200.8 \\
DMFT(G$_{0}$)  & 1.5 $\muB$& 402.1 & 1.4 $\muB$ & 188.8 \\
DMFT(G$_{\text{HF}}$)  & 1.6 $\muB$& 402.3 & 1.5 $\muB$ & 185.0 \\
LDA+$U$(AMF)   & 2.0 $\muB$& 403.6 & 2.0 $\muB$ & 201.2 \\
LDA+$U$(FLL)   & 2.0 $\muB$& 403.7 & 2.0 $\muB$ & 202.6 \\
\hline
Exp. 1~\cite{Bushmeleva:2006jn} & $1.63(6)$  & $408.582(7)$  & - & n/a  \\
Exp. 2~\cite{Hamlin:2007ct} &  n/a & $409.31(8)$ & - & $192(3)$\\
Exp. 3~\cite{Kanbayasi:1976kn} &  $1.57(13)$ & $ 406.5(18) $ & - & n/a\\
\hline
\hline 
\multicolumn{5}{|c|}{Cubic structure}\\

%\hline
%AM05   & 1.395 $\muB$ & 415.9 & 1.462 $\muB$  &   172.7   \\    
%PBE96  & 1.806 $\muB$ & 435.5 & 2.062 $\muB$  &   152.0   \\    
%PW91   & 1.324 $\muB$ & 402.2 & 1.226 $\muB$  &   191.3   \\    
%PZ81   & 1.345 $\muB$ & 402.9 & 1.234 $\muB$  &   198.9   \\    
%vBH72  & 1.334 $\muB$ & 403.9 & 1.409 $\muB$  &   196.6   \\
%\hline

\hline
AM05   & 1.4 $\muB$ & 415.9 & 1.5 $\muB$  &   172.7   \\    
PBE96  & 1.8 $\muB$ & 435.5 & 2.0 $\muB$  &   152.0   \\    
PW91   & 1.3 $\muB$ & 402.2 & 1.2 $\muB$  &   191.3   \\    
PZ81   & 1.3 $\muB$ & 402.9 & 1.2 $\muB$  &   198.9   \\    
vBH72  & 1.3 $\muB$ & 403.9 & 1.4 $\muB$  &   196.6   \\
\hline
\end{tabular}
% rspt110K  & 1.6 $\mu_{B}$& 402.1 & 1.4 $\mu_{B}$ & 188.9 \\
\end{center}
\caption{\label{tab:fig1}{ Equilibrium volume, bulk modulus and magnetic moment per formula unit calculated with different methods, for the orthorhombic and cubic phases of \SRO. For the LDA+$U$ and LDA+DMFT calculations a $U=3 \text{ eV}$ was used. The magnetic moment is reported both for the calculated equilibrium volume and for the experimental volume of Bushmeleva {\itshape et al.} \cite{Bushmeleva:2006jn} } }%The evolution of the moment with respect to $U$ is discussed in detail in section \ref{sec:elcorr}. The delicate nature of the magnetic moment is evident, the parametrization by Hedin and Lundquist results in a moment almost 30\% larger than that of the parametrization by Perdew and Zunger. Note also that the magnetic moment obtained with PBE96 is already in the higher end of the scale at the experimental volume, and even becomes larger at equilibrium volume. The experimental bulk modulus B0 is $\approx$192GPa \cite{Hamlin:2007ct}. The experimental volume is $\approx$408 a.u.$^{3}$}.}
\label{table1}
\end{table}
%
%In Fig.~\ref{fig:fig1}    
% The values of the volumes obtained  trend is displayed in Fig.~\ref{fig:fig1}, and detailed information about equilibrium volume etc. in Table \ref{tab:fig1}. 

We begin by examining the volumes for the orthorhombic structure given by the different methods. The calculated equilibrium volume is similar for all three LDA functionals, only a slight over binding is seen for this compound (around 1.6\%-2\%), while LDA+DMFT expands the lattice slightly resulting in only a 1.5\% underestimation relative to the experimental volume. 
There is a prominent difference between the equilibrium volumes calculated with the LDA (PW91, PZ81, vBH72) and GGA (PBE96) functionals. Typically an LDA treatment results in over binding, while GGA in under binding and an overestimation of the ground state volume. 
%The differences between LDA and GGA is most prominent when calculating equilibrium volumes, typically LDA results in over binding, and GGA under binding and an overestimation of the ground state volume. 
%
 As expected the PBE96 functional under binds also for \SRO, resulting in a 6\% larger volume than the experimental value. On the other hand, the AM05 functional is in better agreement with the experimental volume with only about 1\% under binding. 
 With the exception of PBE96, which leads to a strong overestimation of the volume, all other functionals considered in this study reproduce rather well the experimental volume. Regarding the bulk modulus, the experimental values are in good agreement with the calculated values for all the functionals used with the exception of PBE96 and AM05 (see Table~\ref{table1}). 
% Of the functionals considered here, they all reproduce the observed values for the volume, perhaps with the exception of PBE96 which results in too large volume. Similar conclusions can be made for the bulk modulus, where PBE96 and maybe AM05 stand out to reproduce experiments with lowest accuracy for this material. The other method reproduce the bulk modulus with good accuracy.

In order to estimate the effect of the octahedral tiltings and rotations on the properties of this compound, we have also considered the cubic phase of \SRO. The calculated equilibrium volume and bulk modulus through different methods are reported in Table~\ref{table1} and Fig.~\ref{fig:fig1}. The differences in bulk modulus and equilibrium volumes between the cubic and orthorhombic phases illustrate the importance of the octahedral distortions for the cohesive properties of \SRO.

%\textit{Comparing the cubic and orthorhombic structures it is evident that the bulk modulus is slightly lower in the orthorhombic phase and the equilibrium volume is somewhat bigger in the cubic phase.} %magnetic moment is not as sensitive to volume changes. 
%This illustrates the importance of the octahedral distortions for the cohesive properties of \SRO. %\textit{Elaborate on the importance of the rotations}

% The experimental bulk modulus B0 is $\approx$192GPa \cite{Hamlin:2007ct}. The experimental volume is $\approx$408 a.u.$^{3}$.

\subsection{Calculated spin moments}
\begin{figure}[htbp]
\begin{center}
\includegraphics[width=0.544\textwidth,clip]{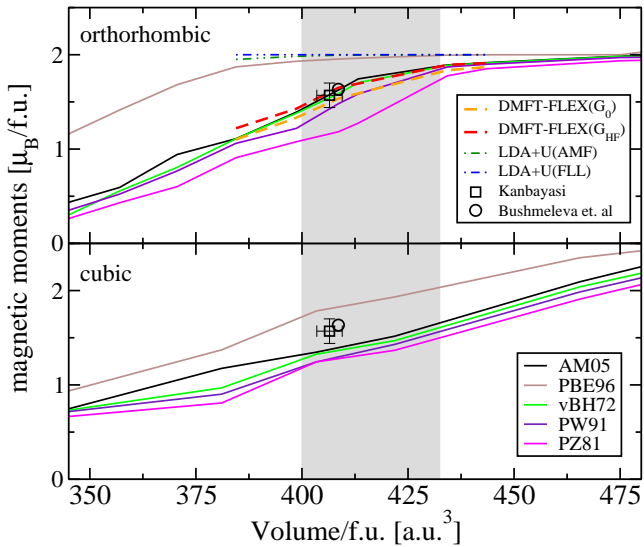} \\
\vspace{-0.5cm} 
\caption{(colour online) Magnetic moment variations induced by changes in volume for both the orthorhombic (upper panel) and the cubic (lower panel) phases. For each chosen value of the volume, the moments have been calculated with different LDA/GGA functionals, LDA+DMFT and LDA+$U$ (here we show only the results for $U=3 \text{ eV}$). The shaded area marks the range of values for the calculated equilibrium volumes. Experimental data-points (including the reported error bars) for the saturation moment for single crystal bulk samples of the orthorhombic phase are included, originating from Refs.~\onlinecite{Kanbayasi:1976kn} (square) and \onlinecite{Bushmeleva:2006jn} (circle). For the latter, the reported error bars are not larger than the symbols we use, and therefore omitted. Finally, notice that the experimental values reported in the lower panel (cubic phase) come also from the orthorhombic phase, and represent only a guide for the eye to facilitate the comparison with the upper panel (see main text).} % We also included the error bars for the experimental data, as reported in Refs.
\label{fig:fig2}
\end{center}
\end{figure}

In Fig.~\ref{fig:fig2} we report the variations of magnetic moment induced by changes in volume and structural properties, as predicted by various computational methods. Before discussing the observed trends, we should mention that we deal with \textit{isobaric} volume changes, meaning that the $a$ to $b$ to $c$ ratio has been conserved for all calculations.%, thus no strain effects have been considered. 
 
%
%\begin{figure}[htbp]
%\begin{center}
%\begin{tabular}{c}
%\includegraphics[width=0.5\textwidth,clip]{fig1a.pdf} \\
%\vspace{-0.75cm} 
%\includegraphics[width=0.5\textwidth,clip]{fig1b.pdf} \\
%\end{tabular}
%\caption{Energy vs volume for a) the orthorhombic structure and b) the cubic structure for different LDA/GGA functionals, LDA+DMFT and LDA+U, with $U$=3.0 eV.  Moment vs volume for b) the orthorhombic structure and d) the cubic structure for different LDA/GGA functionals. The experimental volume, 408 a.u.$^{3}$, is plotted with a dotted line. Experimental data-points for saturation volume for single crystal bulk samples are included, originating from Refs \onlinecite{Kanbayasi:1976fg} and \onlinecite{Bushmeleva:2006jn}.}
%\label{fig:fig1}
%\end{center}
%\end{figure}
%
%a fact which highlights the need to treat the electron-electron interactions with care. Many theoretical studies focuses on the experimental volume, here we highlight differences in the magnetic properties of \SRO  for the experimental volume with different treatment of the exchange correlation within DFT. 

In order to account for possible basis-set dependence of the calculated values, we have also investigated how the trend in volume changes as a function of the chosen basis set, i.e. the difference between a triple-$\kappa$ basis and a double $\kappa$-basis. We find that for the equilibrium volume the differences are minute, $\approx 0.1 \%$ for the cubic structure and $\approx 0.001 \%$ for the orthorhombic structure for both LDA (vBH72) and GGA (AM05). 
For the magnetic moment at the experimental volume we find that the cubic phase is insensitive to whether we use a double- or triple-$\kappa$ basis set, with difference of $\approx 0.5 \%$ for both LDA and GGA. The orthorhombic structure is more sensitive, with a $\approx 5 \%$ difference in moment, highlighting the importance of a good basis set in the interstitial region to describe the effects of the octahedral distortions.

%The reason why different LDA functionals lead to such different moments, is expected to be primarily due to the partly localized character of the Ru $4d$-electrons, which are the main origin of the magnetic moment. 

We highlight the variations of the magnetic properties of SrRuO$_3$ due to different treatments of the exchange correlation within DFT as well as DFT+DMFT (see Table~\ref{tab:fig1}  and Fig.~\ref{fig:fig2}). These differences are much larger than those that one normally finds in other compounds. %Here we highlight differences in the magnetic properties of \SRO for the experimental volume with different treatment of the exchange correlation within DFT
In the orthorhombic structure, the delicate nature of the magnetic moment is evident, the parametrization by von Barth and Hedin %and Lundquist 
results in a moment almost 30\% larger than that of the parametrization by Perdew and Zunger. Note also that the magnetic moment obtained with PBE96 is already in the higher end of the scale at the experimental volume, and even becomes larger at the calculated equilibrium volume.

 For the orthorhombic structure, the most accurate predictions of the magnetic moments for the experimental volumes have been provided by the LDA+DMFT-SPTF (G$_{HF}$), by the LDA (vBH72) and the GGA functional, AM05. The LDA+$U$, for both double counting corrections (AMF and FLL), as well as the PBE96 provide an overestimated magnetic moment for the entire volume range considered, while providing a very robust magnetic moment with respect to volume changes for almost the entire range considered. The LDA+DMFT-SPTF (G$_0$) only slightly underestimates the value of the moment (Fig~\ref{fig:fig2}). When varying the volume, the changes in the magnetic moment follow the same trend (with less than $7 \%$ difference in values) when treated with LDA+DMFT-SPTF (G$_0$) or LDA+DMFT-SPTF (G$_{HF}$). The volume (for both orthorhombic and cubic phases) has been varied well below and above the calculated equilibrium values, in order to offer a complete picture of the magnetic moments sensitivity to \textit{isobaric} volume changes. In the orthorhombic structure, at volumes larger than \V, the calculated magnetic moments converge to the same value independently of the methods used, while for volumes smaller than \V the magnetic moments are much more sensitive to the different methods employed. The grey shaded area in Fig.~\ref{fig:fig2} marks the range of calculated (with different methods) equilibrium volumes. Within this range (that encompasses $35 \text{ a.u.}^3$ around the experimental volume) for a volume compression of $8 \%$ the decrease registered in the magnetic moments vary from $\approx 20 \%$ for LDA+DMFT and vBH72 to almost $40 \%$ for PZ81. In the lower panel of Fig.~\ref{fig:fig2}, we chose to compare the calculated properties for the cubic phase of \SRO with the experimental values obtained for the orthorhombic phase as well. The reason is that we performed calculations for the cubic structure to better assess for the impact of large volume variations on the rotations and tilting of the oxygen octahedra, and not to investigate the physical properties of the high temperature cubic phase. Such an analysis is beyond the scope of this work.  
 
The magnetic moment as a function of volume shows similar trends for all the functionals, as long as we are far from a half-metallic state, the key issue lies in the displacement of the curves. %In fact, $\frac{\Delta M}{\Delta V}\simeq0.001$, i.e. a change in volume of 1\% induces a change in moment of about $0.1\muB$ for all functionals. 
However PBE96 localizes the electrons too much (discussed in details below) and this produces a magnetic moment that is insensitive to volume changes around the experimental volume. Gu {\itshape et al.}~\cite{Gu:2012uq} reports on the change in moment for a \SRO film relaxed on a SrTiO$_{3}$ substrate, showing that a strain of 1\% gives a change in moment of the order of $0.1~\muB$. The relaxation procedure is however not volume conserving and volume changes might be partly responsible for this. 
Around the experimental volume, both LDA+DMFT methods, with SPTF in G$_0$ and G$_{HF}$, show the same behaviour of magnetic moments with changes in volume.
%The effect of strain on the moment is Grutter {\itshape et al.}~\cite{Grutter:2012bp} argues that the strain must reach a certain level, where the crystal fields are mixed enough to allow for a high spin solution of 4$\muB$ per Ru atom. 
%\textit{Considering the $d$-occupation of the Ru ions  it appears to be similar to Cr, where the chemistry of the material is such that the moment is very sensitive to changes in volume \comment{citation needed, Olle? Can we make some more quantitative statement, e.g. the density of wave function nodes, ELF, coop etc?}.} 
Fig.~\ref{fig:fig2} shows that one should be careful when comparing calculated magnetic moments, because they can differ strongly with respect to the considered volume and the method used.
%The magnetic moment at the equilibrium volume is therefore rather difficult to calculate accurately. 
The experimental situation is such that the strong moment anisotropy makes it difficult to achieve saturation moments for polycrystalline samples. However, a few single crystal results have been reported (e.g. Ref.~\onlinecite{Kanbayasi:1976kn, Takizawa:2005vn,Toyota:2005ge}).
%\textit{
As suggested in Ref.~\onlinecite{Rondinelli:2008jy}, %one could in principle achieve a better theoretical estimate of the magnetic moment if one applies 
on-site correlations to the Ru 4d states improve the description of the electronic structure. In general, smaller changes in the description of the electronic structure, as given by different parametrizations of LDA, can cause (after the self-consistent cycle is done) variations in the calculated magnetic moment. With respect to this quantity the AM05 functional performs excellently. The LDA+U method creates a too strong localization of the $d$-electrons (discussed below), and the magnetic moment overshoots the experimental one. The equilibrium volume also increases with $U$, an effect of the decreased contribution to bonding by the increasingly localized $d$-states. The LDA+DMFT method includes dynamic screening, reducing the tendency of localisation. In addition, this method increases the electron effective mass. For the chosen value of $U$, i.e. $3 \text{ eV}$, the magnetic properties of \SRO are very well reproduced by theory (see Fig.~\ref{fig:fig2}). %a more elaborate review of the LDA+U and LDA+DMFT results are found in section \ref{sec:elcorr}.

\subsection{Effects of on-site Coulomb interaction}\label{sec:elcorr}
\begin{figure}[b]
\begin{center}
\includegraphics[width=0.5\textwidth,clip]{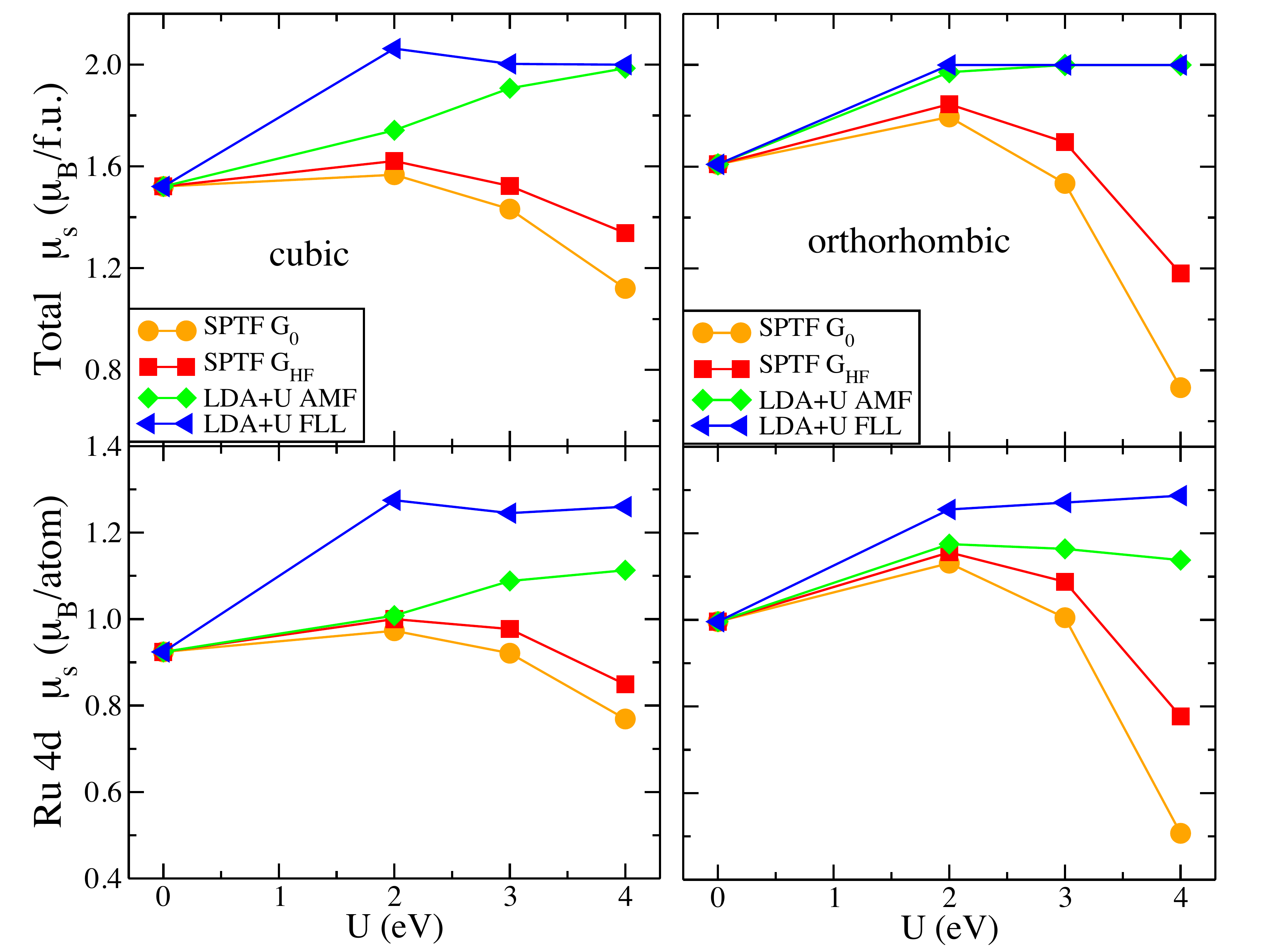}
\caption{(colour online) Calculated total magnetic moments and the Ru-4d moments, for both cubic and orthorhombic phases, as a function of the Hubbard $U$, when using the solvers SPTF(G$_{0}$) and SPTF(G$_{\mathrm{HF}}$) within LDA+DMFT and different double-counting corrections (AMF and FLL) within LDA+$U$.}
\label{fig:fig3}
\end{center}
\end{figure}

%\comment{Cut from Grutter et al: It is, at first, surprising that the films with the highest resistivity also have the highest carrier concentration. How- ever, (111) films also exhibit significantly decreased mobility that we believe to be associated with a higher incidence of defects and increased mosaic spread, as observed in residual resistivity and x-ray diffraction measurements, respectively. So, the enhancement of the carrier concentration, and, hence, the density of states at the Fermi level can be correlated with the magnetism. }
Due to the delicate nature of the electronic structure of \SRO, it becomes of interest to investigate how the strength of the electron correlations influences the electronic structure and magnetic properties of this material. We therefore inspect the evolution of the magnetic moment when changing the strength of the on-site Coulomb interaction, in both the cubic and orthorhombic phases (Fig.~\ref{fig:fig3}). 
%We argue that the starting LDA functional is arbitrary, as the magnetism arises from the Ru $4d$-shell, where we apply the on-site Coulomb interaction. 
%The evolution of the moment with respect to $U$ is discussed in detail in section \ref{sec:elcorr}. 
%
%\textit{The delicate nature of the magnetic moment is evident, the parametrization by Hedin and Lundquist results in a moment almost 30\% larger than that of the parametrization by Perdew and Zunger. Note also that the magnetic moment obtained with PBE96 is already in the higher end of the scale at the experimental volume, and even becomes larger at equilibrium volume.}
%
Using LDA+$U$ with AMF double-counting correction the moment of the orthorhombic phase increases monotonically from the von Barth-Hedin LDA value of $1.6~\muB$ towards $2~\muB$ (see Fig.~\ref{fig:fig3}). This means that we quickly reach (already for $U=2 \text{ eV}$) a higher moment than the highest reported experimental one. The high moment solution, characterised by an integer moment of $2~\muB$, is due to the fact that the electronic structure becomes half-metallic, which we will notice in the density of states below. Also, a further analysis in connection to the LDA+$U$ (FLL double-counting correction) calculation results in a higher moment with increasing values of $U$. When we consider \SRO in the cubic structure, the moment reaches its maximum for $U=2 \text{ eV}$, approaching the (AMF) solution at $U=4 \text{ eV}$. For the orthorhombic phase, there are only very small differences between the values obtained by employing the different double counting corrections on LDA+$U$. In the cubic phase the differences are bigger.

%
%\begin{figure}[htbp]
%\begin{center}
%\begin{tabular}{cc}
%\includegraphics[width=0.27\textwidth,clip]{fig3a.pdf} &
% \hspace{-0.2cm}\includegraphics[width=0.24\textwidth,clip]{fig3b.pdf} \\
%\end{tabular}
%\caption{(colour online) Difference in the calculated total magnetic moments and the Ru \textit{4d} moments, for both cubic and orthorhombic phases, as a function of the Hubbard U, when using the solvers SPTF(G$^{0}$) and SPTF(G$^{\mathrm{HF}}$) within DMFT and different double-counting corrections (AMF and FLL) within LDA+U.}
%\label{fig:fig3}
%\end{center}
%\end{figure}
As the electrons in \SRO also have an itinerant character, it is a good idea to use a method that allows for dynamic screening of the coulomb interaction, such as LDA+DMFT.  %, as more itinerant electrons (that contribute to a large part of the binding) are treated within the vBH72 scheme.
The LDA+DMFT calculation with the SPTF(G$_{0}$) solver results in an initial slight increase, followed by a strong reduction of the magnetic moment when $U$ increases, for both the cubic phase as well as the orthorhombic one. In the cubic phase the moment is always lower or equal to the LDA value, for all values of $U$ considered. Electron correlations, i.e. effects beyond Hartree-Fock theory, are often argued to cause a reduction in the exchange splitting, and hence a lowered magnetic moment. The cubic phase of SrRuO$_3$ seems to follow this rule. For the orthorhombic phase the scenario is a little bit more complex, since initially there is an increase in the moment with increasing $U$, after which a strong decrease is found (Fig.~\ref{fig:fig3}). 
Calculations based on SPTF(G$_{\mathrm{HF}}$) and SPTF(G$_{0}$) give similar results, although when treated within SPTF(G$_{\mathrm{HF}}$) the moment is not reduced as strongly, with increasing $U$. %new below
The reason behind the decrease in moment is that the main peaks of predominant Ru $t_{2g}$ character in both spin-channels are pushed symmetrically towards the Fermi level, as can be seen in Fig.~\ref{fig:fig6}. The narrowing of the spectra and subsequent development of satellite peaks is a known phenomena best displayed in single band models \cite{Kotliar2006}. Also, with increasing $U$ the adequacy of the SPTF solver decreases due to the known under screening of the coulomb interaction, partly cured in the SPTF(G$_{\mathrm{HF}}$) formulation, and the inability of the solver to describe the formation of local moments.

%{\bf explicatie}
%The hybridization with oxygen has been fully taken into account by the charge-self-consistent cycle. \\
%
%\vspace{-0.5cm}

\begin{figure}[t]
\begin{center}
\includegraphics[width=0.7\textwidth,clip]{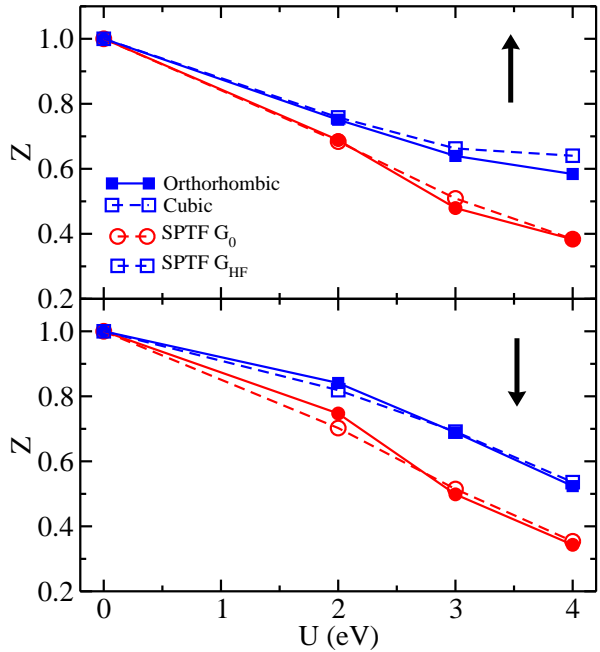}
\caption{(colour online) Dependence of the calculated quasi particle weight $Z$ on the Hubbard $U$ and the different LDA+DMFT solvers SPTF(G$_{0}$) (circles) and SPTF(G$_{\mathrm{HF}}$) (squares). The filled symbols represent the orthorhombic phase, while the empty symbols represent the cubic phase. The two panels give spin resolved information.
\label{fig:fig4}}
\end{center}
\end{figure}

In order to get a deeper insight into the electronic structure of \SRO, we investigate the sensitivity of the effective mass to the considered on-site Coulomb interactions. The only method in this study capable of capturing a renormalised quasiparticle weight $Z$ is LDA+DMFT. The experimental mass enhancement arises both from electron-electron interaction as well as electron-boson interactions, as recently suggested by Shai {\itshape et al.}~\cite{Shai2013} The latter can be electron-phonon or electron-magnon interactions. We calculate the spin resolved contribution from the electron-electron interaction only. For both the cubic and orthorhombic phases, the dependence on the Hubbard $U$ of the spin resolved quasiparticle weight values averaged over the full $d$-shell is depicted in Fig.~\ref{fig:fig4}.

Experimental data for the effective mass (inverse of $Z$) are rather abundant. EELS data by Cox {\itshape et al.}~\cite{Cox:1983wx} show an effective mass of about ${m^{*}}/{m_{0}} \approx 3.6$, this relies on the assumption that only one free electron is available per Ru atom. On the other hand, angular resolved de Haas van Alphen  data indicate at least seven bands crossing the Fermi level, with an effective mass ratio of $4.1 \leq {m^{*}}/{m_{0}} \leq 6.9$.~\cite{Alexander:2005iu} Specific heat measurements~\cite{Cao:1997id,Ahn:1999ks} show ${m^{*}}/{m_{0}} \approx 3.0-3.1$, while older experiments by Allen {\itshape et al.}~\cite{Allen1996} and Okamoto {\itshape et al.}~\cite{Okamoto:1999wk} provide a value closer to 4.43. This leads to a quasiparticle weight in the vicinity of $0.15 < Z < 0.35$. We draw the attention of the reader to the fact that the experimental values corresponding to the mass enhancements that we mentioned above, have been extracted by different methods from the raw experimental data. In some cases even values calculated from first principles have been used in the fitting equations. To some extent, these values might depend on the way they have been extracted from the measured spectra. 

We consider the EELS and specific heat measurements to be most representative when comparing to the orbitally averaged quasiparticle weight. The orbital average using SPTF(G$_{0}$) for $U=4 \text{ eV}$ compare well with the experimental number, with $Z \approx 0.4$, considering the quenching of the magnetic moment this is however not representative for the compound. Overall values for $U=3 \text{ eV}$ are more reasonable, and therefore we estimate the orbital average $Z \approx 0.58$ for the electronic part of the mass enhancement in SPTF(G$_{0}$). Projecting on local $t_{2g}$ and $e_{g}$ manifolds and averaging over spin we get $Z_{t_{2g}} \approx 0.45$ and $Z_{e_{g}} \approx 0.79$. This is well in line with a previous study for a spin-degenerate $t_{2g}$ manifold by means of the Continuous-time Quantum Monte Carlo (CT-QMC) solver~\cite{deMedici:2011uq}, where a value $Z \approx 0.4$ was obtained for a band filling similar to \SRO. Due to the under screening of the Coulomb interaction in the bare SPTF solver, one should obtain a better estimate through SPTF(G$_{\mathrm{HF}}$), which gives a slightly higher $Z \approx 0.66$. Therefore, we estimate the local electronic correlation to give an orbital averaged contribution of $0.4 < Z < 0.7$, somewhat underestimated with respect to the experimental values reported above. 
However, we have strong reasons to believe that a significant part of the experimental mass enhancement comes from electron-boson coupling. In fact, recent ARPES measurements by Shai {\itshape et al.}~\cite{Shai2013} reveal a kink in the spectra, in the vicinity of the Fermi level, which is a text-book example of electron-boson coupling, most likely magnons. This contribution is not considered in the single-site LDA+DMFT approach, which includes uniquely electronic contributions. Another possibility is, instead, the presence of a crossover to a competing energy scale, as seen in SrVO$_{3}$ \cite{Nekrasov:2006ha}. This energy scale can be shown to be related to the effective Kondo energy scale of the interacting lattice\cite{Held:2013dq}, and is likely beyond the capabilities of the perturbative impurity solver employed by us. Addressing the kink in the quasiparticle spectra rigorously would require the use of a non-perturbative impurity solver and the inclusion of non-local correlation effects in a single computational scheme.

%\comment{Here, before discussing the SOC, we could mention that have contributions from all orbitals, but t2g states may dominate m*. In this case we should mention that we agree well with Antoine.} 

Another possible reason behind the difference between theory and experiments could be that the inclusion of spin-orbit coupling has a large impact on the quasiparticle weights; it has previously been shown that spin-orbit coupling is very important for the geometry of the Fermi surface of the related compound Sr$_{2}$RhO$_{4}$.~\cite{Liu:2008io} However, in our calculations, we find only a small difference in the calculated quasiparticle weights, with and without the inclusion of spin-orbit coupling effects, amounting to 0.76\%. Notice that effects due to the spin-orbit coupling are considered both in the electronic structure part, as well as in the solver.

%Another possibility is that the electron-phonon contribution is unusually large, though substantial contributions are expected to be captured by the specific heat measurements by the associated $T^{3}$ behaviour.

%\comment{Comment on non T$^{2}$ behavior of the resistivity, cite Georges. Comment on high temperature local moments and DMFT, transverse spin-fluctuations using ASD, cite Igors Mn paper, ASD paper. DMFT (or at least DLM) nessesary for the high temperature properties of a material which retains it's local moment above Tc.}

\subsection{Spectral properties}
There are two experimental studies of the valence band spectrum of orthorhombic SrRuO$_3$ that we are aware of. One is by D. Toyota {\itshape et al.}~\cite{Toyota:2005ge} and represents an investigation of a single crystal film, with a thickness up to 100 monolayers (ML). The other study has been performed on a polycrystalline bulk sample.~\cite{Okamoto:1999wk} We chose to compare our calculations to the single crystal data, which are expected to better represent a calculation which does not consider effects of grain boundaries.

\begin{figure}[b]
\begin{center}
\includegraphics[width=0.53\textwidth]{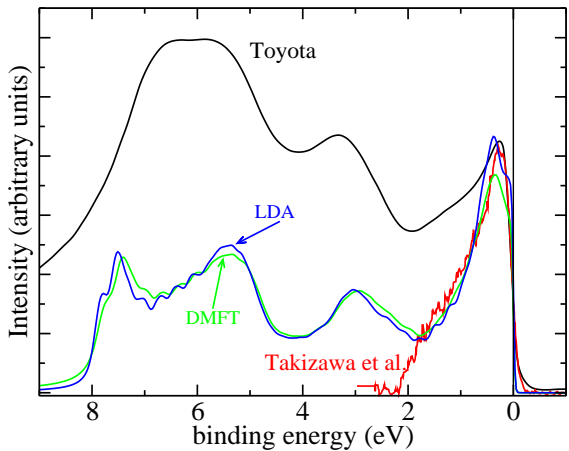}
\caption{(colour online) Photoemission spectra of bulk \SRO from Ref.~\onlinecite{Takizawa:2005vn}  (red line) and 100 monolayers thick \SRO films~\cite{Toyota:2005ge} (black line) compared to calculated spectra obtained within LDA (blue line) and LDA+DMFT, SPTF-G$_{HF}$ with $U=2 \text{ eV}$ (green line). Both PES were taken at 600~eV. Note that the intensity is given in arbitrary units for both theory and experiment. The theoretical spectra is calculated by summing the products of the $l$-projected DOS with the $l$-projected cross sections for the photoelectron excitation with photons of 600~eV energy from Ref.~\onlinecite{Yeh:1985vt}. No additional broadening of the theoretical spectra is added.}
\label{fig:fig5}
\end{center}
\end{figure}

\begin{figure}[th]
\begin{center}
\hspace{-0.3cm}\includegraphics[width=0.67\textwidth,angle=90,clip]{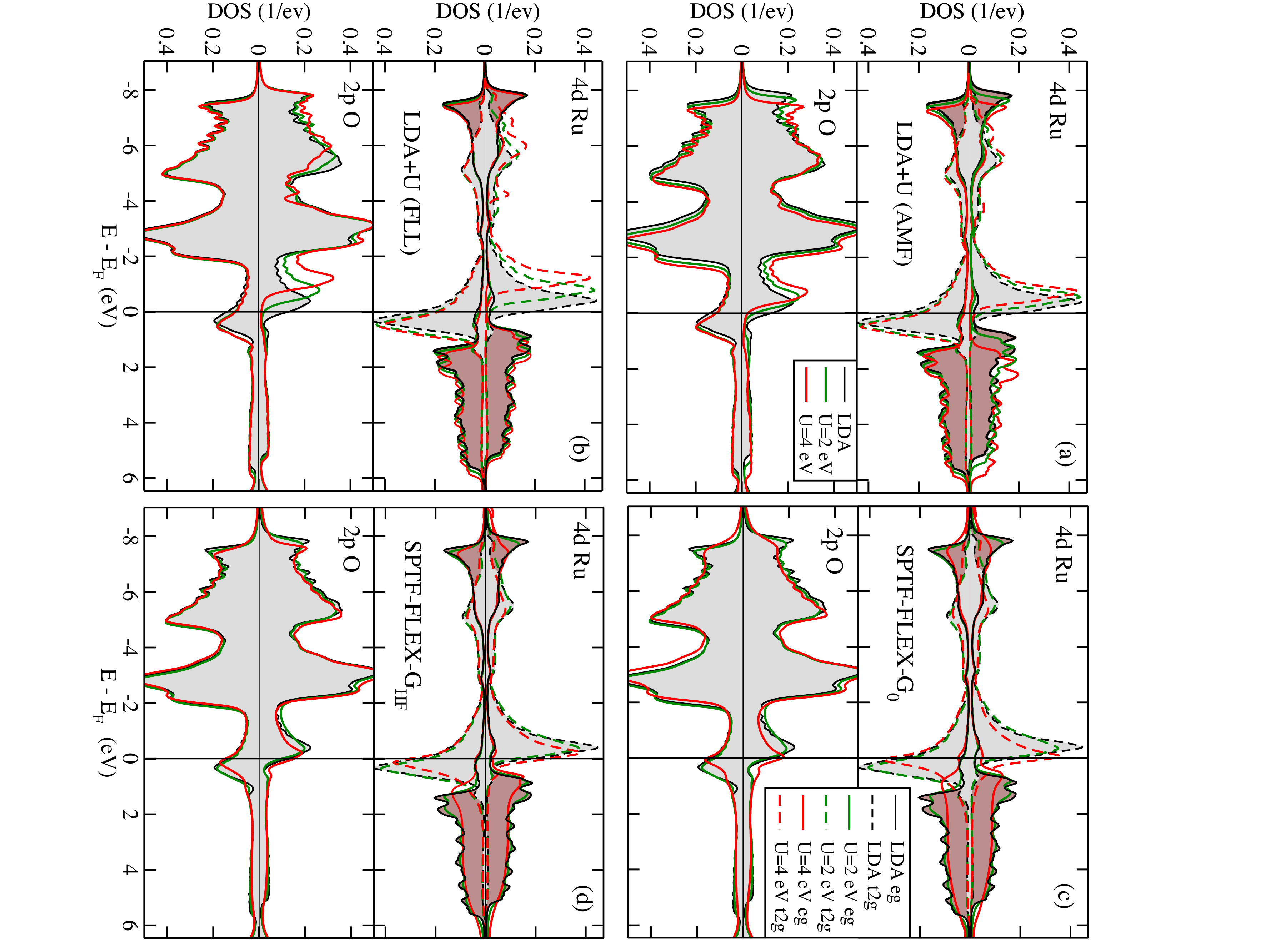}
\vspace{-0.6cm}
\caption{%\comment{How shoudl we do the PES comparison? Using the LDA as reference is clear, as that is the reference in the other plots. If we have the 4d and the total in the same graph it is harder to see that we actually have a substantial 4d-character at the peaks. Note also that the axis on the older DOS plots should be scaled up by a factor of 4.}
(colour online) Calculated DOS for \SRO in the orthorhombic structure. Left-panel: LDA and LDA+$U$ results with different double counting corrections: (a) around mean field (AMF) and (b) fully localized limit (FLL). Right-panel: calculated DOS within LDA+DMFT with two different solvers, (c) SPTF-G$_0$ and (d) SPTF-G$_{HF}$.}
\label{fig:fig6}
\end{center}
\end{figure}

The photoemission data by Toyota {\itshape et al.} for 100 ML \SRO show three very clear peaks (Fig.~\ref{fig:fig5} shows this experimental spectrum). The highest peak is located 0.3-0.5~eV below the Fermi level. In addition there is one level found at approximately $E_F -3.5$~eV and a larger, broader feature between $4$ and $8$~eV below the Fermi level. These two lower-most peaks are mainly composed of oxygen $2p$-states, though the lowest lying peak has a considerable amount of Ru 4d-character. The peak close to the Fermi level is almost entirely of Ru 4d character, and is therefore most suitable to compare with and to determine the quality of our calculated spectral functions, shown in Fig.~\ref{fig:fig5} and Fig.~\ref{fig:fig6}. As stressed above, all spectral properties were calculated for $U=2$, 3 and 4 eV, but for improving the readability of the figure we chose not to present the spectra for $U=3 \text{ eV}$. These are situated, as expected, in between the results for $U=2 \text{ eV}$ and $U=4 \text{ eV}$ for all the investigated techniques. 
%%%%%%%  GAAH
The LDA+$U$ result, displayed in Fig.~\ref{fig:fig6}, shows a half-metallic state for $U>3 \text{ eV}$ similar to what Jeng {\itshape et al.}~\cite{Jeng:2006iq} obtained. This is not consistent with experimental findings.~\cite{Bushmeleva:2006jn} Both LDA and LDA+DMFT provide metallic states in both spin channels. In fact the LDA and the LDA+DMFT spectra are surprisingly similar. The top most peak shows some renormalization and is pushed closer to the Fermi level in the LDA+DMFT scheme. Comparing to previously published spectral functions with similar methodology we see an overall agreement.~\cite{Jakobi:2011cs} Both  LDA and  LDA+DMFT methodologies show good agreement with experimental photo-emission data. 
In Fig~\ref{fig:fig5} we compare our theoretical curves (LDA and LDA+DMFT results) to the experimental data of Toyota {\itshape et al.} and Takizawa {\itshape et al.}~\cite{Toyota:2005ge,Takizawa:2005vn} The theoretical spectra are calculated by summing the products of the $l$-projected DOS with the $l$-projected cross sections for the photoelectron excitation with photons of 600 eV energy from Ref.~\onlinecite{Yeh:1985vt}. In the 0-2 eV binding energy regime, both experiments as well as both theoretical curves are similar. A peak appearing around 3 eV binding energy is well reproduced by theory, it is mainly of oxygen character with some Ru 4d hybridisation. 
In the experiments by Toyota {\itshape et al.} a large feature of hybridising O 2p and Ru 4d states is found between 5 and 8 eV binging energy, also this is well captured by theory, although slightly broader. The relative intensities of all peaks are in good agreement with the experimental spectra.
% Although the Ru \textit{4d} band-width differs for the LDA and LDA+DMFT spectra. This is a probable reason why the mass enhancement differs by $\sim$30\% (Fig.~\ref{fig:fig4}).

As a final note to the comparison with experimental data performed here and in the previous subsection, we should mention that the LDA+DMFT approach has some restrictions when applied to real systems.~\cite{Vollhardt:2012fk,Kotliar2006} To make an example the choice of the local orbitals is far from unambiguous, especially for delocalized states such as Ru-4d electrons. Of particular importance for our analysis is the mapping into an effective model with a static Coulomb interaction $U$. In fact, it was recently found~\cite{PhysRevB.85.035115} that in principle one can not reach a good agreement for both mass enhancement and spectra at the same time with a static $U$. 

%%\hspace{-1cm}
%\begin{figure}[htbp]\centering
%\begin{tabular}{cc}
%\includegraphics[width=0.26\textwidth,clip]{fig5a_1.pdf} &
%\hspace{-0.3cm} \includegraphics[width=0.2264\textwidth,clip]{fig5c.pdf} \\
%\includegraphics[width=0.26\textwidth,clip]{fig5b.pdf} &
%\hspace{-0.2cm}\includegraphics[width=0.234\textwidth,clip]{fig5d.pdf} \\
%%\includegraphics[width=0.27\textwidth,clip]{PES_Ru4d_tot.pdf} &
%%\hspace{-0.73cm} \includegraphics[width=0.27\textwidth,clip]{PES_DFTinone.pdf} \\
%%%\multicolumn{2}{l}{\includegraphics[width=0.40\textwidth,clip]{PES_Ru4d_tot.pdf}}\\
%\end{tabular}
%\caption{%\comment{How shoudl we do the PES comparison? Using the LDA as reference is clear, as that is the reference in the other plots. If we have the 4d and the total in the same graph it is harder to see that we actually have a substantial 4d-character at the peaks. Note also that the axis on the older DOS plots should be scaled up by a factor of 4.}
%(colour online) Calculated DOS for \SRO in the orthorhombic structure. Left-panel: within LDA and LDA+U with different double counting corrections: (a) around mean field (AMF) and (b) fully localized limit (FLL). Right-panel: calculated DOS within DMFT with two different solvers, (c) SPTF-FLEX G$^0$ and (d) SPTF-FLEX G$^{HF}$.}
%\label{fig:fig5}
%\end{figure}

\subsection{Effects of spin-orbit coupling}
Spin-orbit interaction introduces a coupling between spin-space and real-space, giving rise to a number of interesting phenomena. For example the magneto-crystalline anisotropy (MCA) which is important from a technological perspective, as it dictates the ability of the moment to remain in the same direction under an external field. \SRO is regarded to have a very strong MCA for being a transition metal compound. It is also argued to have a universal MCA, i.e. the easy axis is always aligned to the strain field.~\cite{Grutter:2012bp} 
%\comment{MAE vs volume is still cooking, initial results point towards that Grutter's analysis is correct in the interval of approx. 400-430 a.u.3 volume. In that case strain is the completely dominant reason for "universal magnetic anisotropy" for this volume range. } 

% (120nm; after 80nm it is reported that the strain field is relaxed) Cropped out, ppl should read the ref.
Grutter {\itshape et al.}~\cite{Grutter:2012bp}  determined the orbital moments for thick films to be in the order of $0.06~\muB$. Strained films reach much higher orbital moments as the crystal field splitting is further reduced, up to $0.3~\muB$ is reported.~\cite{Grutter:2012bp} The calculated orbital moment, obtained here, for the bulk orthorhombic structure at the equilibrium volume is around $0.01~\muB$ for all functionals, i.e. lower than the reported experimental value.  

% \comment{High quality single crystal films reported in 26 Q. X. Jia, F. Chu, C. D. Adams, X. D. Wu, M. Hawley, J. H. Cho, A. T. Findikoglu, S. R. Foltyn, J. L. Smith, and T. E. Mitchell, J. Mater. Res. 11, 2263 (1996). 27 W. Siemons, G. Koster, A. Vailionis, H. Yamamoto, D. H. A. Blank, and M. R. Beasley, Phys. Rev. B 76, 075126 (2007). }

\begin{figure}[htbp]
\begin{center}
\includegraphics[width=0.5\textwidth,clip]{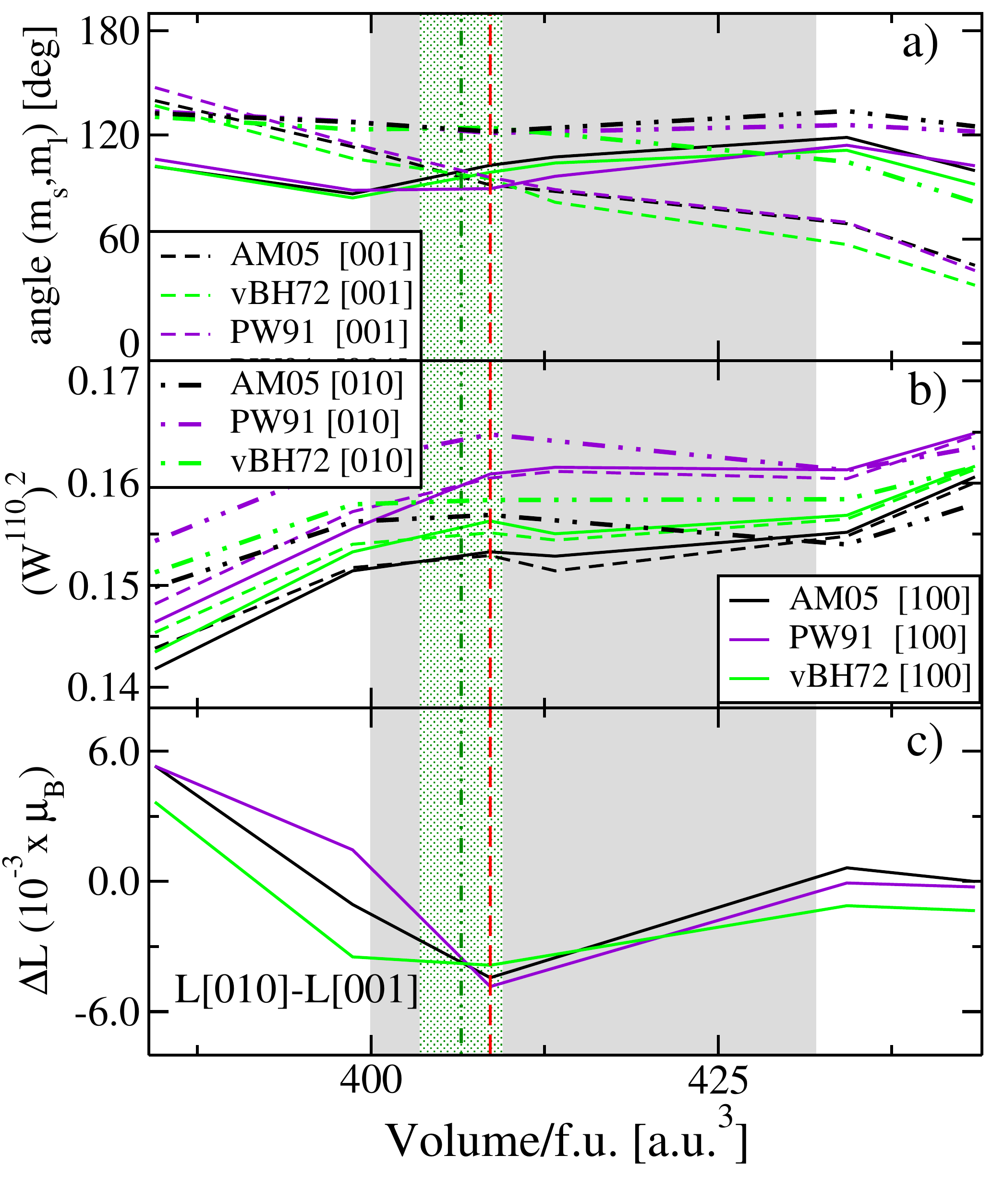}
\caption{(colour online) a) Volume dependence of the angle between the spin and the orbital moment. The angles between the spin and orbital moments have been calculated with the spin-axis aligned along three different crystallographic directions: [100] (solid lines), [010] (dot-dashed lines) and [001] (dashed lines).  b) Square of the $w^{110}$ tensor component, proportional to the isotropic spin-orbit coupling. In (c) the orbital moment anisotropy (OMA) is shown. The experimental volumes value are marked by vertical lines: red dashed lines~\cite{Bushmeleva:2006jn} and green dash-dotted lines.~\cite{Kanbayasi:1976kn} The green shaded area represents the range of the reported experimental errors,~\cite{Kanbayasi:1976kn} while the gray shaded area marks the region of the calculated equilibrium volumes.   
}
\label{fig:fig7}
\end{center}
\end{figure}

\begin{figure}[htbp]
\begin{center}
\includegraphics[width=0.5\textwidth,clip]{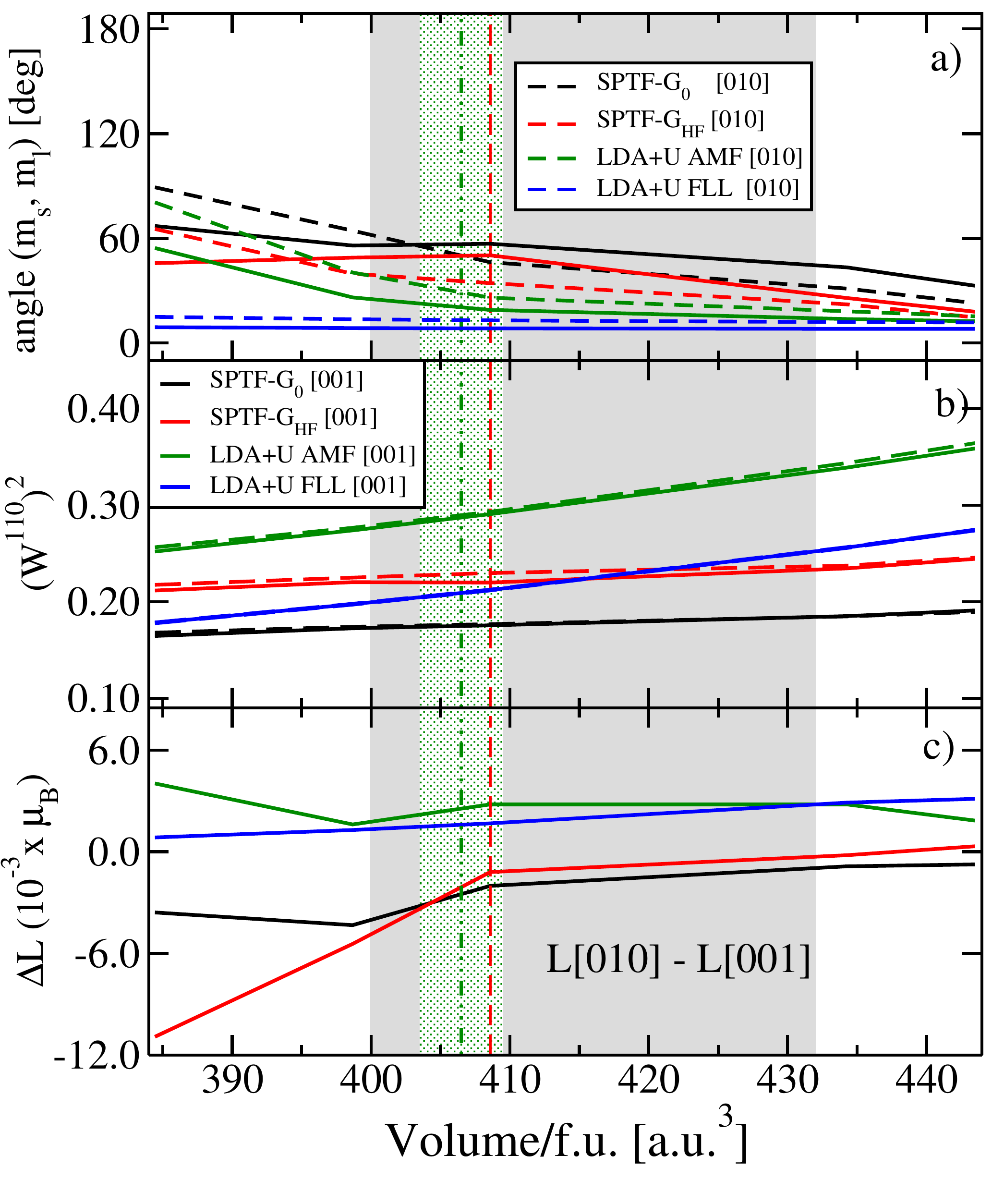}
\caption{(colour online) As for Fig.~\ref{fig:fig7} using different approaches for adding on-site orbital dependent coulomb interaction within: LDA+DMFT with SPTF-G$_0$, LDA+DMFT with SPTF-G$_{HF}$, LDA+$U$ in AMF and LDA+$U$ in FLL).}
\label{fig:fig8}
\end{center}
\end{figure}

More interestingly, the orbital moment is not collinear with respect to the spin moment for the orthorhombic structure. This severely complicates measurements of magneto-optical properties,~\cite{Solovyev:1998ec} and XMCD measurements of orbital moments, as the sum-rules are expressed in terms of the $z$-components of the spin and orbital moments.~\cite{Thole:1992tm} Experimental measurements of the tilted orbital moments might be possible using EMCD,~\cite{Rusz:2011cf} where atomic resolution might be accomplished using electron vortex beams.~\cite{Verbeeck:2010is}

The non-collinear ordering of spin and orbital moments in distorted perovskites is intrinsically coupled to the tiltings and rotations of the oxygen octahedra, as previously noted by Refs.~\onlinecite{Treves:1962cs,Solovyev:1997bt}. 
%When an external magnetic field is used to rotate the magnetic moments away from the easy axis, it is known that the non-collinear ordering between the spin and orbital moments appears. 

In surface geometries where we have a strong out of plane MCA, it is known that the presence of an in-plane magnetic field, aligning the spin-moment in the plane, results in a non-collinear ordering of spin and orbital moments. In these geometries the strong MCA will try to rotate the orbital moments towards the spin moment, while a strong crystal field counteracts the alignment to the spins, and induces an orbital moment component along the easy axis.~\cite{Durr:1996he} Thus, it may result that the spin and orbital moments are not always aligned. 

In distorted perovskites this picture becomes more complicated. The strong exchange coupling drives the spin moments to be in principle collinear,~\cite{elk-calcs} but the rotated oxygen octahedra results in crystal fields that induce orbital components in other directions than parallel to the spin. The orbital moment components aligned orthogonal to the spin moment are antiferromagnetically aligned with respect to each other, whereas the parallel component is ferromagnetically aligned. The ordering type is presented in Table~\ref{tab:AFMordering} for all considered functionals, including the LDA+$U$ and LDA+DMFT calculations. 

The ordering of the different components of the orbital moment is completely insensitive to the choice of the exchange-correlation functional and additional treatment of correlation in terms of LDA+$U$ and LDA+DMFT (in both approaches one has to take care that a rotationally invariant description of the Coulomb interaction is used). It is also insensitive to volume changes and is a direct effect of the distortions of the oxygen octahedra, since the tilting and rotation angles are taken from experimental data only the strength of the crystal field and hybridisation will change with changing volume. As Table~\ref{tab:AFMordering} shows, volume changes under such conditions do not affect the type of ordering, however, the magnitude of the components of the orbital moment do change. This causes the angle between spin and orbital moment to have a volume dependence even without changes in the tilting and rotation angles of the oxygen octahedra.

\begin{table}[b]
\begin{center}
\begin{tabular}{|c|c|c|c|}
\hline
  & \multicolumn{3}{c|}{Ordering of orbital components}  \\   
\hline
Spin alignment  & m$_x$ & m$_y$ & m$_z$ \\ 
\hline
[010]  & AFM C-type & Ferromagnetic & AFM G-type \\   
 
[001]  & AFM A-type & AFM G-type & Ferromagnetic \\    
\hline
\end{tabular}
\end{center}
\caption{\label{tab:AFMordering}{The table shows the ordering between the $x-$, $y-$ and $z-$ components of the orbital moments in the four different Ru octahedra for the spin aligned along the easy [010] axis or [001] hard axis. This ordering is the same for all methods tested, including LDA+$U$ and LDA+DMFT, and is robust with respect to changes in volume.}}
% Should we comment on the rotated structure wrt the cif?
\end{table}
% Should we be consistent in convention and call Ferro AFM B-type? Confusing notation according to me, but that's how it is... I prefer Ferro

The magnitude of each component of the orbital moment is controlled by the strength of the crystal field and hybridisation. The hybridisation changes significantly between different treatments of exchange-correlation, hence the angle between the spin and orbital moments is an extremely sensitive probe for differences in the theoretical treatment. Investigating this we see a clear trend when treating the system within LDA/GGA (Fig.~\ref{fig:fig7}) and within LDA+$U$/DMFT (Fig.~\ref{fig:fig8}). When the electrons in \SRO are treated within LDA/GGA, the angles between the spin and orbital moments at the experimental volume have almost the same values ($\approx 90-100^\circ$) for spin-axis along the [010] or [001] directions, and about 125 $^\circ$ for the [010] direction. For the [010] or [001] directions the moments go from being mostly anti-parallel as dictated by Hund's third rule, to being mostly parallel when increasing volume. With spin quantization axis along [010] the system remains mostly anti-parallel for all volumes, as can be seen in panel a) of Fig.~\ref{fig:fig7}. When \SRO is treated within LDA+$U$ and LDA+DMFT, the angles between moments exhibit almost the same trend for both [001] and [010] spin directions, for the whole range of considered volumes. Independent of the functionals used for treating the electronic system, for very large volumes the spin and orbital moments tend to align parallel to each other, see panel a) of Fig.~\ref{fig:fig8}). At the experimental volume, the angles determined within LDA+$U$ and LDA+DMFT show a stronger variation than when treated with LDA/GGA functionals, leading to variations in direction from 8$^{\circ}$ to 57$^\circ$. The magnitude of the angles is also here determined by the interplay between the crystal fields and the hybridisation, it is evident that the different treatments of exchange and correlation on the Ru 4d-shell impacts the hybridisation significantly. The impact of the spin-orbit coupling on the energetics can be estimated by calculating the expectation value of $l\cdot s$. Using the tensor moment formalism~\cite{Bultmark:2009gm} the isotropic spin-orbit coupling is equivalent to the $w^{110}=(ls)^{-2}\sum_{i}\vec{l}_{i}\cdot\vec{s}_{i}$ tensor. This is plotted in panel b) of Fig.~\ref{fig:fig7} and ~\ref{fig:fig8} as a function of volume. One can see a clear increase in the role of the spin-orbit coupling on the energetics as the volume increases. This indicates the dominance of the kinetic hybridization/crystal field effects for smaller volumes.  The $w^{110}$ component is largest for the [010] direction, indicating that the system gains the most energy from the spin-orbit coupling when the spin-axis is aligned in this direction, as expected for the easy-axis.

%%%   Rewrite by Oscar 20131031
%%% New from Oscar 2013-11-08
For conventional monoatomic solids the MCA is proportional to the anisotropy of the orbital moment.~\cite{Bruno:1989th} Due to the non-trivial angle between the spin and orbital moment and the polyatomic nature of the compound this is not expected to hold for \SRO. The orbital magnetic anisotropy  presented here isolates effects from changes in volume, as the tilting and rotation angles are kept fixed. 
The OMA is expected to change, as well as the MCA, when lattice relaxations are taken into account. However, the tilting and rotation angles have been shown not to change significantly under volume changes of a few percent.~\cite{Zayak2006} Uniaxial strain, as occurring in thin films, is expected to give a very large contribution as the oxygen octahedra will be elongated/compressed in the direction of the strain, resulting in a more anisotropic environment for the Ru-ion. 
%%% New from Oscar 2013-11-08
% \begin{itemize}
% \item add $\Delta$ L [010]-[001] to Fig. 7 ...for AM05, PW91, vBH72
% \item check the hybridisation function \\
% - difference between the integrated hybrid. funct. for the cubic and orthorhombic phases \\
% - for the orthorhombic phase, diff. in the projected hybrid. funct. per non-equiv. Ru atom 
% \item explain the S,L angles
% \end{itemize}

As mentioned above, we have estimated the spin-orbit coupling influence on the mass enhancement to be negligible. %(isn't this due to the fact that we only have contributions from the el-el int.; probably electron-magnon is much higher ... etc )

\section{Conclusions}
We have investigated the electronic structure, cohesion and magnetism of SrRuO$_3$, both in the high temperature cubic phase as well as in the low temperature orthorhombic phase. Several parametrizations of the LDA and GGA functionals were used, as well as the LDA+$U$ method and LDA+DMFT in combination with the FLEX-derived solver SPTF. We find that LDA, AM05, LDA+$U$ and LDA+DMFT with $U=3 \text{ eV}$ reproduce the equilibrium volume and bulk modulus with reasonable accuracy, whereas the GGA functional PBE96 seems to be less accurate. When it comes to the magnetism, LDA, AM05 and LDA+DMFT reproduce the delicate spin moments of the orthorhombic structure with good accuracy, whereas LDA+$U$ and PBE96 overestimate the spin moments significantly. Taking all this information into account, we conclude that LDA, AM05 and LDA+DMFT reproduce the ground state properties of SrRuO3 very well, if one considers $U=3 \text{ eV}$. This conclusion is obviously dependent on the exact value of the parameter $U$, but does not change drastically if one consider values between 2 eV and 4 eV. For the latter, however, a significant quenching of the magnetic moment is observed in LDA+DMFT when using SPTF as impurity solver. LDA+DMFT provides also information about the mass enhancement due to local electronic correlations. Our analysis points to that SrRuO$_3$ is a weakly correlated metal, for which the effective mass is enhanced significantly compared to that given by a single particle theory. The LDA+$U$ results  are found to localize the 4d states of the Ru atom too strongly for realistic values of $U$, in line with the experience with self-interaction corrected theories.~\cite{Etz:2012uq} We also find that the calculated spin moments are quite sensitive to the equilibrium volume, a result which should be possible to verify experimentally. We show furthermore that the coupling between spin and orbital moments is highly non-collinear, with angles ranging from 0 to about 150 degrees, depending on the level of accuracy for the exchange correlation functional and the volume considered for the calculation. An experimental verification of this behaviour would be highly interesting, although challenging due to the symmetry of the orthorhombic crystal structure. Regarding the electronic structure we find that both LDA, AM05 and LDA+DMFT reproduce the measured valence band spectrum with good accuracy. This result is independent on the exact value of $U$ in the range 2-4 eV, given that the differences in spectral properties are more or less within the range of the experimental accuracy. Therefore, in conclusion, all our results point to GGA (AM05) or LDA+DMFT as methods of choice for further investigations on \SRO. However, one should consider that the Hubbard parameter $U$, whose ab-initio determination is still not feasible at a high level of accuracy, introduces an additional uncertainty in the LDA+DMFT approach, which makes this method preferred only if one is interested in fine details of the spectral properties such as the mass enhancement.

 %One feature is missing in the theoretical description and that is the distinct peak at 3 eV binding energy, as shown in the experiment of Toyota {\itshape et al.}53 We speculate that this feature may represent a satellite feature which is too weak in the DMFT+FLEX theory, similar to the valence band satellite of Ni. It is however also possible that this experimental peak is caused by defects like O vacancies or by surface effects. 

%\textit{Our contribution to the scientific community interested in the correct description of the physics of strontium ruthenates is a ...complex/complete/exhaustive..... study based on dynamical mean-field theory. In this theoretical investigation, besides describing the electronic and magnetic properties of \SRO within DFT+DMFT (considering a full charge self consistency), we have also constantly compared these results with calculations based on LDA/GGA and LDA+U. In order the be able to make an assessment of the most suitable method for an accurate description of \SRO, we related our calculated quantities to experimental photoemission spectra, mass enhancements and magnetic moments. %The DMFT calculation used the spin-polarized T-matrix fluctuation exchange solver. }
%We find that overall both the GGA by Armiento and Mattsson and the LDA+DMFT reproduce rather well experimentally measured quantities: both in what concerns the magnetic moments as well as details %of the photoelectron spectra. We would recommend these two methods for further investigations on the \SRO compound. 

\begin{acknowledgments}
The support from the Swedish Research Council (VR) is thankfully acknowledged. O.G. is thankful for support by the eSSENCE network for funding time to improve implementations that allowed for this study. O.E. acknowledges support from the KAW foundation and the ERC (Grant No. 247062-ASD). The computer calculations have been performed at the Swedish high performance centres HPC2N, NSC and UPPMAX under grants provided by the Swedish National Infrastructure for Computing (SNIC). 
\end{acknowledgments}

\bibliography{SrRuO3,extra,all}

\begin{thebibliography}{69}%
\makeatletter
\providecommand \@ifxundefined [1]{%
 \@ifx{#1\undefined}
}%
\providecommand \@ifnum [1]{%
 \ifnum #1\expandafter \@firstoftwo
 \else \expandafter \@secondoftwo
 \fi
}%
\providecommand \@ifx [1]{%
 \ifx #1\expandafter \@firstoftwo
 \else \expandafter \@secondoftwo
 \fi
}%
\providecommand \natexlab [1]{#1}%
\providecommand \enquote  [1]{``#1''}%
\providecommand \bibnamefont  [1]{#1}%
\providecommand \bibfnamefont [1]{#1}%
\providecommand \citenamefont [1]{#1}%
\providecommand \href@noop [0]{\@secondoftwo}%
\providecommand \href [0]{\begingroup \@sanitize@url \@href}%
\providecommand \@href[1]{\@@startlink{#1}\@@href}%
\providecommand \@@href[1]{\endgroup#1\@@endlink}%
\providecommand \@sanitize@url [0]{\catcode `\\12\catcode `\$12\catcode
  `\&12\catcode `\#12\catcode `\^12\catcode `\_12\catcode `\%12\relax}%
\providecommand \@@startlink[1]{}%
\providecommand \@@endlink[0]{}%
\providecommand \url  [0]{\begingroup\@sanitize@url \@url }%
\providecommand \@url [1]{\endgroup\@href {#1}{\urlprefix }}%
\providecommand \urlprefix  [0]{URL }%
\providecommand \Eprint [0]{\href }%
\providecommand \doibase [0]{http://dx.doi.org/}%
\providecommand \selectlanguage [0]{\@gobble}%
\providecommand \bibinfo  [0]{\@secondoftwo}%
\providecommand \bibfield  [0]{\@secondoftwo}%
\providecommand \translation [1]{[#1]}%
\providecommand \BibitemOpen [0]{}%
\providecommand \bibitemStop [0]{}%
\providecommand \bibitemNoStop [0]{.\EOS\space}%
\providecommand \EOS [0]{\spacefactor3000\relax}%
\providecommand \BibitemShut  [1]{\csname bibitem#1\endcsname}%
\let\auto@bib@innerbib\@empty
%</preamble>
\bibitem [{\citenamefont {Lee}\ \emph {et~al.}(2009)\citenamefont {Lee},
  \citenamefont {Allan}, \citenamefont {Wang}, \citenamefont {Farrell},
  \citenamefont {Grigera}, \citenamefont {Baumberger}, \citenamefont {Davis},\
  and\ \citenamefont {Mackenzie}}]{Lee:2009hm}%
  \BibitemOpen
  \bibfield  {author} {\bibinfo {author} {\bibfnamefont {J.}~\bibnamefont
  {Lee}}, \bibinfo {author} {\bibfnamefont {M.~P.}\ \bibnamefont {Allan}},
  \bibinfo {author} {\bibfnamefont {M.~A.}\ \bibnamefont {Wang}}, \bibinfo
  {author} {\bibfnamefont {J.}~\bibnamefont {Farrell}}, \bibinfo {author}
  {\bibfnamefont {S.~A.}\ \bibnamefont {Grigera}}, \bibinfo {author}
  {\bibfnamefont {F.}~\bibnamefont {Baumberger}}, \bibinfo {author}
  {\bibfnamefont {J.~C.}\ \bibnamefont {Davis}}, \ and\ \bibinfo {author}
  {\bibfnamefont {A.~P.}\ \bibnamefont {Mackenzie}},\ }\href@noop {} {\bibfield
   {journal} {\bibinfo  {journal} {Nature Physics}\ } (\bibinfo {year}
  {2009})}\BibitemShut {NoStop}%
\bibitem [{\citenamefont {Maeno}\ \emph {et~al.}(1994)\citenamefont {Maeno},
  \citenamefont {Hashimoto}, \citenamefont {Yoshida}, \citenamefont
  {Nishizaki}, \citenamefont {Fujita}, \citenamefont {Bednorz},\ and\
  \citenamefont {Lichtenberg}}]{Maeno1994}%
  \BibitemOpen
  \bibfield  {author} {\bibinfo {author} {\bibfnamefont {Y.}~\bibnamefont
  {Maeno}}, \bibinfo {author} {\bibfnamefont {H.}~\bibnamefont {Hashimoto}},
  \bibinfo {author} {\bibfnamefont {K.}~\bibnamefont {Yoshida}}, \bibinfo
  {author} {\bibfnamefont {S.}~\bibnamefont {Nishizaki}}, \bibinfo {author}
  {\bibfnamefont {T.}~\bibnamefont {Fujita}}, \bibinfo {author} {\bibfnamefont
  {J.}~\bibnamefont {Bednorz}}, \ and\ \bibinfo {author} {\bibfnamefont
  {F.}~\bibnamefont {Lichtenberg}},\ }\href@noop {} {\bibfield  {journal}
  {\bibinfo  {journal} {Nature}\ }\textbf {\bibinfo {volume} {372}},\ \bibinfo
  {pages} {532} (\bibinfo {year} {1994})}\BibitemShut {NoStop}%
\bibitem [{\citenamefont {Mravlje}\ \emph {et~al.}(2011)\citenamefont
  {Mravlje}, \citenamefont {Aichhorn}, \citenamefont {Miyake}, \citenamefont
  {Haule}, \citenamefont {Kotliar},\ and\ \citenamefont
  {Georges}}]{Mravlje:2011gk}%
  \BibitemOpen
  \bibfield  {author} {\bibinfo {author} {\bibfnamefont {J.}~\bibnamefont
  {Mravlje}}, \bibinfo {author} {\bibfnamefont {M.}~\bibnamefont {Aichhorn}},
  \bibinfo {author} {\bibfnamefont {T.}~\bibnamefont {Miyake}}, \bibinfo
  {author} {\bibfnamefont {K.}~\bibnamefont {Haule}}, \bibinfo {author}
  {\bibfnamefont {G.}~\bibnamefont {Kotliar}}, \ and\ \bibinfo {author}
  {\bibfnamefont {A.}~\bibnamefont {Georges}},\ }\href@noop {} {\bibfield
  {journal} {\bibinfo  {journal} {Physical Review Letters}\ }\textbf {\bibinfo {volume}
  {106}},\ \bibinfo {pages} {096401}  
  (\bibinfo {year} {2011})}\BibitemShut {NoStop}%
\bibitem [{\citenamefont {Klein}\ \emph {et~al.}(1999)\citenamefont {Klein},
  \citenamefont {Antognazza}, \citenamefont {Geballe}, \citenamefont
  {Beasley},\ and\ \citenamefont {Kapitulnik}}]{Klein:1999tz}%
  \BibitemOpen
  \bibfield  {author} {\bibinfo {author} {\bibfnamefont {L.}~\bibnamefont
  {Klein}}, \bibinfo {author} {\bibfnamefont {L.}~\bibnamefont {Antognazza}},
  \bibinfo {author} {\bibfnamefont {T.~H.}~\bibnamefont {Geballe}}, \bibinfo
  {author} {\bibfnamefont {M.~R.}~\bibnamefont {Beasley}}, \ and\ \bibinfo
  {author} {\bibfnamefont {A.}~\bibnamefont {Kapitulnik}},\ }\href@noop {}
  {\bibfield  {journal} {\bibinfo  {journal} {Physical Review B}\ }\textbf
  {\bibinfo {volume} {60}},\ \bibinfo {pages} {1448} (\bibinfo {year}
  {1999})}\BibitemShut {NoStop}%
\bibitem [{\citenamefont {Mangalam}\ and\ \citenamefont
  {Sundaresan}(2009)}]{Mangalam:2009dc}%
  \BibitemOpen
  \bibfield  {author} {\bibinfo {author} {\bibfnamefont {R.~V.~K.}\
  \bibnamefont {Mangalam}}\ and\ \bibinfo {author} {\bibfnamefont
  {A.}~\bibnamefont {Sundaresan}},\ }\href@noop {} {\bibfield  {journal}
  {\bibinfo  {journal} {Materials Research Bulletin}\ }\textbf {\bibinfo
  {volume} {44}},\ \bibinfo {pages} {576} (\bibinfo {year} {2009})}\BibitemShut
  {NoStop}%
\bibitem [{\citenamefont {Mazin}\ and\ \citenamefont
  {Singh}(1997)}]{Mazin:1997es}%
  \BibitemOpen
  \bibfield  {author} {\bibinfo {author} {\bibfnamefont {I.~I.}~\bibnamefont
  {Mazin}}\ and\ \bibinfo {author} {\bibfnamefont {D.~J.}~\bibnamefont {Singh}},\
  }\href@noop {} {\bibfield  {journal} {\bibinfo  {journal} {Physical Review
  B}\ }\textbf {\bibinfo {volume} {56}},\ \bibinfo {pages} {2556} (\bibinfo
  {year} {1997})}\BibitemShut {NoStop}%
\bibitem [{\citenamefont {Dabrowski}\ \emph {et~al.}(2004)\citenamefont
  {Dabrowski}, \citenamefont {Chmaissem}, \citenamefont {Klamut}, \citenamefont
  {Kolesnik}, \citenamefont {Maxwell}, \citenamefont {Mais}, \citenamefont
  {Ito}, \citenamefont {Armstrong}, \citenamefont {Jorgensen},\ and\
  \citenamefont {Short}}]{Dabrowski:2004ix}%
  \BibitemOpen
  \bibfield  {author} {\bibinfo {author} {\bibfnamefont {B.}~\bibnamefont
  {Dabrowski}}, \bibinfo {author} {\bibfnamefont {O.}~\bibnamefont
  {Chmaissem}}, \bibinfo {author} {\bibfnamefont {P.~W.}~\bibnamefont {Klamut}},
  \bibinfo {author} {\bibfnamefont {S.}~\bibnamefont {Kolesnik}}, \bibinfo
  {author} {\bibfnamefont {M.}~\bibnamefont {Maxwell}}, \bibinfo {author}
  {\bibfnamefont {J.}~\bibnamefont {Mais}}, \bibinfo {author} {\bibfnamefont
  {Y.}~\bibnamefont {Ito}}, \bibinfo {author} {\bibfnamefont {B.~D.}~\bibnamefont
  {Armstrong}}, \bibinfo {author} {\bibfnamefont {J.~D.}~\bibnamefont
  {Jorgensen}}, \ and\ \bibinfo {author} {\bibfnamefont {S.}~\bibnamefont
  {Short}},\ }\href@noop {} {\bibfield  {journal} {\bibinfo  {journal}
  {Physical Review B}\ }\textbf {\bibinfo {volume} {70}} 
  ,\ \bibinfo {pages} {014423} (\bibinfo {year}
  {2004})}\BibitemShut {NoStop}%
\bibitem [{\citenamefont {Kanbayasi}(1976)}]{Kanbayasi:1976kn}%
  \BibitemOpen
  \bibfield  {author} {\bibinfo {author} {\bibfnamefont {A.}~\bibnamefont
  {Kanbayasi}},\ }\href@noop {} {\bibfield  {journal} {\bibinfo  {journal} 
  {Journal of the Physical Society of Japan}\ }\textbf {\bibinfo {volume} 
  {41}},\ \bibinfo {pages} {1876}
  (\bibinfo {year} {1976})}\BibitemShut {NoStop}%
\bibitem [{\citenamefont {Koster}\ \emph {et~al.}(2012)\citenamefont {Koster},
  \citenamefont {Klein}, \citenamefont {Siemons}, \citenamefont {Rijnders},
  \citenamefont {Dodge}, \citenamefont {Eom}, \citenamefont {Blank},\ and\
  \citenamefont {Beasley}}]{Koster:2012gr}%
  \BibitemOpen
  \bibfield  {author} {\bibinfo {author} {\bibfnamefont {G.}~\bibnamefont
  {Koster}}, \bibinfo {author} {\bibfnamefont {L.}~\bibnamefont {Klein}},
  \bibinfo {author} {\bibfnamefont {W.}~\bibnamefont {Siemons}}, \bibinfo
  {author} {\bibfnamefont {G.}~\bibnamefont {Rijnders}}, \bibinfo {author}
  {\bibfnamefont {J.~S.}\ \bibnamefont {Dodge}}, \bibinfo {author}
  {\bibfnamefont {C.-B.}\ \bibnamefont {Eom}}, \bibinfo {author} {\bibfnamefont
  {D.~H.~A.}\ \bibnamefont {Blank}}, \ and\ \bibinfo {author} {\bibfnamefont
  {M.~R.}\ \bibnamefont {Beasley}},\ }\href@noop {} {\bibfield  {journal}
  {\bibinfo  {journal} {Review of Modern Physics}\ }\textbf {\bibinfo {volume} {84}},\
  \bibinfo {pages} {253} (\bibinfo {year} {2012})}\BibitemShut {NoStop}%
\bibitem [{\citenamefont {Ziese}\ \emph {et~al.}(2010)\citenamefont {Ziese},
  \citenamefont {Vrejoiu}, \citenamefont {Pippel}, \citenamefont {Esquinazi},
  \citenamefont {Hesse}, \citenamefont {Etz}, \citenamefont {Henk},
  \citenamefont {Ernst}, \citenamefont {Maznichenko}, \citenamefont {Hergert},\
  and\ \citenamefont {Mertig}}]{Ziese:2010fk}%
  \BibitemOpen
  \bibfield  {author} {\bibinfo {author} {\bibfnamefont {M.}~\bibnamefont
  {Ziese}}, \bibinfo {author} {\bibfnamefont {I.}~\bibnamefont {Vrejoiu}},
  \bibinfo {author} {\bibfnamefont {E.}~\bibnamefont {Pippel}}, \bibinfo
  {author} {\bibfnamefont {P.}~\bibnamefont {Esquinazi}}, \bibinfo {author}
  {\bibfnamefont {D.}~\bibnamefont {Hesse}}, \bibinfo {author} {\bibfnamefont
  {C.}~\bibnamefont {Etz}}, \bibinfo {author} {\bibfnamefont {J.}~\bibnamefont
  {Henk}}, \bibinfo {author} {\bibfnamefont {A.}~\bibnamefont {Ernst}},
  \bibinfo {author} {\bibfnamefont {I.~V.}\ \bibnamefont {Maznichenko}},
  \bibinfo {author} {\bibfnamefont {W.}~\bibnamefont {Hergert}}, \ and\
  \bibinfo {author} {\bibfnamefont {I.}~\bibnamefont {Mertig}},\ }\href
  {http://link.aps.org/doi/10.1103/PhysRevLett.104.167203} {\bibfield
  {journal} {\bibinfo  {journal} {Physical Review Letters}\ }\textbf {\bibinfo
  {volume} {104}},\ \bibinfo {pages} {167203} (\bibinfo {year}
  {2010})}\BibitemShut {NoStop}%
\bibitem [{\citenamefont {Fang}\ \emph {et~al.}(2003)\citenamefont {Fang},
  \citenamefont {Nagaosa}, \citenamefont {Takahashi}, \citenamefont {Asamitsu},
  \citenamefont {Mathieu}, \citenamefont {Ogasawara}, \citenamefont {Yamada},
  \citenamefont {Kawasaki}, \citenamefont {Tokura},\ and\ \citenamefont
  {Terakura}}]{Fang:2003uq}%
  \BibitemOpen
  \bibfield  {author} {\bibinfo {author} {\bibfnamefont {Z.}~\bibnamefont
  {Fang}}, \bibinfo {author} {\bibfnamefont {N.}~\bibnamefont {Nagaosa}},
  \bibinfo {author} {\bibfnamefont {K.~S.}\ \bibnamefont {Takahashi}}, \bibinfo
  {author} {\bibfnamefont {A.}~\bibnamefont {Asamitsu}}, \bibinfo {author}
  {\bibfnamefont {R.}~\bibnamefont {Mathieu}}, \bibinfo {author} {\bibfnamefont
  {T.}~\bibnamefont {Ogasawara}}, \bibinfo {author} {\bibfnamefont
  {H.}~\bibnamefont {Yamada}}, \bibinfo {author} {\bibfnamefont
  {M.}~\bibnamefont {Kawasaki}}, \bibinfo {author} {\bibfnamefont
  {Y.}~\bibnamefont {Tokura}}, \ and\ \bibinfo {author} {\bibfnamefont
  {K.}~\bibnamefont {Terakura}},\ }\href@noop {} {\bibfield  {journal}
  {\bibinfo  {journal} {Science}\ }\textbf {\bibinfo {volume} {302}},\ \bibinfo
  {pages} {92} (\bibinfo {year} {2003})}\BibitemShut {NoStop}%
\bibitem [{\citenamefont {Felner}\ \emph {et~al.}(2006)\citenamefont {Felner},
  \citenamefont {Nomura},\ and\ \citenamefont {Nowik}}]{Felner:2006hm}%
  \BibitemOpen
  \bibfield  {author} {\bibinfo {author} {\bibfnamefont {I.}~\bibnamefont
  {Felner}}, \bibinfo {author} {\bibfnamefont {K.}~\bibnamefont {Nomura}}, \
  and\ \bibinfo {author} {\bibfnamefont {I.}~\bibnamefont {Nowik}},\
  }\href@noop {} {\bibfield  {journal} {\bibinfo  {journal} {Physical Review
  B}\ }\textbf {\bibinfo {volume} {73}},\ \bibinfo {pages} {064401}
  (\bibinfo {year} {2006})}\BibitemShut
  {NoStop}%
\bibitem [{\citenamefont {Rondinelli}\ \emph {et~al.}(2008)\citenamefont
  {Rondinelli}, \citenamefont {Caffrey}, \citenamefont {Sanvito},\ and\
  \citenamefont {Spaldin}}]{Rondinelli:2008jy}%
  \BibitemOpen
  \bibfield  {author} {\bibinfo {author} {\bibfnamefont {J.~M.}~\bibnamefont
  {Rondinelli}}, \bibinfo {author} {\bibfnamefont {N.~M.}~\bibnamefont {Caffrey}},
  \bibinfo {author} {\bibfnamefont {S.}~\bibnamefont {Sanvito}}, \ and\
  \bibinfo {author} {\bibfnamefont {N.~A.}~\bibnamefont {Spaldin}},\ }\href@noop
  {} {\bibfield  {journal} {\bibinfo  {journal} {Physical Review B}\ }\textbf
  {\bibinfo {volume} {78}},\ \bibinfo {pages} {155107} (\bibinfo {year}
  {2008})}\BibitemShut {NoStop}%
\bibitem [{\citenamefont {Jakobi}\ \emph {et~al.}(2011)\citenamefont {Jakobi},
  \citenamefont {Kanungo}, \citenamefont {Sarkar}, \citenamefont {Schmitt},\
  and\ \citenamefont {Saha-Dasgupta}}]{Jakobi:2011cs}%
  \BibitemOpen
  \bibfield  {author} {\bibinfo {author} {\bibfnamefont {E.}~\bibnamefont
  {Jakobi}}, \bibinfo {author} {\bibfnamefont {S.}~\bibnamefont {Kanungo}},
  \bibinfo {author} {\bibfnamefont {S.}~\bibnamefont {Sarkar}}, \bibinfo
  {author} {\bibfnamefont {S.}~\bibnamefont {Schmitt}}, \ and\ \bibinfo
  {author} {\bibfnamefont {T.}~\bibnamefont {Saha-Dasgupta}},\ }\href@noop {}
  {\bibfield  {journal} {\bibinfo  {journal} {Physical Review B}\ }\textbf
  {\bibinfo {volume} {83}},\ \bibinfo {pages} {041103}  (\bibinfo {year} {2011})}\BibitemShut {NoStop}%
\bibitem [{\citenamefont {Jeng}\ \emph {et~al.}(2006)\citenamefont {Jeng},
  \citenamefont {Lin},\ and\ \citenamefont {Hsue}}]{Jeng:2006iq}%
  \BibitemOpen
  \bibfield  {author} {\bibinfo {author} {\bibfnamefont {H.-T.}\ \bibnamefont
  {Jeng}}, \bibinfo {author} {\bibfnamefont {S.-H.}\ \bibnamefont {Lin}}, \
  and\ \bibinfo {author} {\bibfnamefont {C.-S.}\ \bibnamefont {Hsue}},\
  }\href@noop {} {\bibfield  {journal} {\bibinfo  {journal} 
  {Physical Review Letters}\ }
  \textbf {\bibinfo {volume} {97}},\ \bibinfo {pages} {067002} (\bibinfo {year} {2006})}\BibitemShut
  {NoStop}%
\bibitem [{\citenamefont {Hadipour}\ and\ \citenamefont
  {Akhavan}(2011)}]{Hadipour:2011ew}%
  \BibitemOpen
  \bibfield  {author} {\bibinfo {author} {\bibfnamefont {H.}~\bibnamefont
  {Hadipour}}\ and\ \bibinfo {author} {\bibfnamefont {M.}~\bibnamefont
  {Akhavan}},\ }\href@noop {} {\bibfield  {journal} {\bibinfo  {journal} 
  {European Physical Journal B}\ }
  \textbf {\bibinfo {volume} {84}},\ \bibinfo {pages} {203}
  (\bibinfo {year} {2011})}\BibitemShut {NoStop}%
\bibitem [{\citenamefont {Kostic}\ \emph {et~al.}(1998)\citenamefont {Kostic},
  \citenamefont {Okada}, \citenamefont {Collins}, \citenamefont {Schlesinger},
  \citenamefont {Reiner}, \citenamefont {Klein}, \citenamefont {Kapitulnik},
  \citenamefont {Geballe},\ and\ \citenamefont {Beasley}}]{Kostic1998}%
  \BibitemOpen
  \bibfield  {author} {\bibinfo {author} {\bibfnamefont {P.}~\bibnamefont
  {Kostic}}, \bibinfo {author} {\bibfnamefont {Y.}~\bibnamefont {Okada}},
  \bibinfo {author} {\bibfnamefont {N.~C.}\ \bibnamefont {Collins}}, \bibinfo
  {author} {\bibfnamefont {Z.}~\bibnamefont {Schlesinger}}, \bibinfo {author}
  {\bibfnamefont {J.~W.}\ \bibnamefont {Reiner}}, \bibinfo {author}
  {\bibfnamefont {L.}~\bibnamefont {Klein}}, \bibinfo {author} {\bibfnamefont
  {A.}~\bibnamefont {Kapitulnik}}, \bibinfo {author} {\bibfnamefont {T.~H.}\
  \bibnamefont {Geballe}}, \ and\ \bibinfo {author} {\bibfnamefont {M.~R.}\
  \bibnamefont {Beasley}},\ }\href
  {http://link.aps.org/doi/10.1103/PhysRevLett.81.2498} {\bibfield  {journal}
  {\bibinfo  {journal} {Physical Review Letters}\ }\textbf {\bibinfo {volume} {81}},\
  \bibinfo {pages} {2498} (\bibinfo {year} {1998})}\BibitemShut {NoStop}%
\bibitem [{\citenamefont {Cox}\ \emph {et~al.}(1983)\citenamefont {Cox},
  \citenamefont {Egdell}, \citenamefont {Goodenough}, \citenamefont {Hamnett},\
  and\ \citenamefont {Naish}}]{Cox:1983wx}%
  \BibitemOpen
  \bibfield  {author} {\bibinfo {author} {\bibfnamefont {P.}~\bibnamefont
  {Cox}}, \bibinfo {author} {\bibfnamefont {R.}~\bibnamefont {Egdell}},
  \bibinfo {author} {\bibfnamefont {J.}~\bibnamefont {Goodenough}}, \bibinfo
  {author} {\bibfnamefont {A.}~\bibnamefont {Hamnett}}, \ and\ \bibinfo
  {author} {\bibfnamefont {C.}~\bibnamefont {Naish}},\ }\href@noop {}
  {\bibfield  {journal} {\bibinfo  {journal} 
  {Journal of Physics C: Solid State Physics}\ }\textbf
  {\bibinfo {volume} {16}},\ \bibinfo {pages} {6221} (\bibinfo {year}
  {1983})}\BibitemShut {NoStop}%
\bibitem [{\citenamefont {Cao}\ \emph {et~al.}(1997)\citenamefont {Cao},
  \citenamefont {McCall}, \citenamefont {Shepard}, \citenamefont {Crow},\ and\
  \citenamefont {Guertin}}]{Cao:1997id}%
  \BibitemOpen
  \bibfield  {author} {\bibinfo {author} {\bibfnamefont {G.}~\bibnamefont
  {Cao}}, \bibinfo {author} {\bibfnamefont {S.}~\bibnamefont {McCall}},
  \bibinfo {author} {\bibfnamefont {M.}~\bibnamefont {Shepard}}, \bibinfo
  {author} {\bibfnamefont {J.E.}~\bibnamefont {Crow}}, \ and\ \bibinfo {author}
  {\bibfnamefont {R.~P.}~\bibnamefont {Guertin}},\ }\href@noop {} {\bibfield
  {journal} {\bibinfo  {journal} {Physical Review B}\ }\textbf {\bibinfo
  {volume} {56}},\ \bibinfo {pages} {321} (\bibinfo {year} {1997})}\BibitemShut
  {NoStop}%
\bibitem [{\citenamefont {Ahn}\ \emph {et~al.}(1999)\citenamefont {Ahn},
  \citenamefont {Bak}, \citenamefont {Choi}, \citenamefont {Noh}, \citenamefont
  {Han}, \citenamefont {Bang}, \citenamefont {Cho},\ and\ \citenamefont
  {Jia}}]{Ahn:1999ks}%
  \BibitemOpen
  \bibfield  {author} {\bibinfo {author} {\bibfnamefont {J.~S.}~\bibnamefont
  {Ahn}}, \bibinfo {author} {\bibfnamefont {J.}~\bibnamefont {Bak}}, \bibinfo
  {author} {\bibfnamefont {H.~S.}~\bibnamefont {Choi}}, \bibinfo {author}
  {\bibfnamefont {T.~W.}~\bibnamefont {Noh}}, \bibinfo {author} {\bibfnamefont
  {J.~E.}~\bibnamefont {Han}}, \bibinfo {author} {\bibfnamefont {Y.}~\bibnamefont
  {Bang}}, \bibinfo {author} {\bibfnamefont {J.~H.}~\bibnamefont {Cho}}, \ and\
  \bibinfo {author} {\bibfnamefont {Q.~X.}~\bibnamefont {Jia}},\ }\href@noop {}
  {\bibfield  {journal} {\bibinfo  {journal} {Physcal Review Letters}\ }\textbf
  {\bibinfo {volume} {82}},\ \bibinfo {pages} {5321} (\bibinfo {year}
  {1999})}\BibitemShut {NoStop}%
\bibitem [{\citenamefont {de'Medici}\ \emph {et~al.}(2011)\citenamefont
  {de'Medici}, \citenamefont {Mravlje},\ and\ \citenamefont
  {Georges}}]{deMedici:2011uq}%
  \BibitemOpen
  \bibfield  {author} {\bibinfo {author} {\bibfnamefont {L.}~\bibnamefont
  {de'Medici}}, \bibinfo {author} {\bibfnamefont {J.}~\bibnamefont {Mravlje}},
  \ and\ \bibinfo {author} {\bibfnamefont {A.}~\bibnamefont {Georges}},\ }\href
  {http://link.aps.org/doi/10.1103/PhysRevLett.107.256401} {\bibfield
  {journal} {\bibinfo  {journal} {Physical Review Letters}\ }\textbf {\bibinfo
  {volume} {107}},\ \bibinfo {pages} {256401} (\bibinfo {year}
  {2011})}\BibitemShut {NoStop}%
\bibitem [{\citenamefont {Georges}\ \emph {et~al.}(2013)\citenamefont
  {Georges}, \citenamefont {Medici},\ and\ \citenamefont
  {Mravlje}}]{Georges:2013}%
  \BibitemOpen
  \bibfield  {author} {\bibinfo {author} {\bibfnamefont {A.}~\bibnamefont
  {Georges}}, \bibinfo {author} {\bibfnamefont {L.~d.}\ \bibnamefont {Medici}},
  \ and\ \bibinfo {author} {\bibfnamefont {J.}~\bibnamefont {Mravlje}},\ }\href
  {\doibase 10.1146/annurev-conmatphys-020911-125045} {\bibfield  {journal}
  {\bibinfo  {journal} {Annual Review of Condensed Matter Physics}\ }\textbf
  {\bibinfo {volume} {4}},\ \bibinfo {pages} {137} (\bibinfo {year}
  {2013})}\BibitemShut {NoStop}%
\bibitem [{\citenamefont {Vollhardt}(2012)}]{Vollhardt:2012fk}%
  \BibitemOpen
  \bibfield  {author} {\bibinfo {author} {\bibfnamefont {D.}~\bibnamefont
  {Vollhardt}},\ }\href {\doibase 10.1002/andp.201100250} {\bibfield  {journal}
  {\bibinfo  {journal} {Annalen der Physik}\ }\textbf {\bibinfo {volume}
  {524}},\ \bibinfo {pages} {1} (\bibinfo {year} {2012})}\BibitemShut {NoStop}%
\bibitem [{\citenamefont {Kotliar}\ \emph {et~al.}(2006)\citenamefont
  {Kotliar}, \citenamefont {Savrasov}, \citenamefont {Haule}, \citenamefont
  {Oudovenko}, \citenamefont {Parcollet},\ and\ \citenamefont
  {Marianetti}}]{Kotliar2006}%
  \BibitemOpen
  \bibfield  {author} {\bibinfo {author} {\bibfnamefont {G.}~\bibnamefont
  {Kotliar}}, \bibinfo {author} {\bibfnamefont {S.~Y.}\ \bibnamefont
  {Savrasov}}, \bibinfo {author} {\bibfnamefont {K.}~\bibnamefont {Haule}},
  \bibinfo {author} {\bibfnamefont {V.~S.}\ \bibnamefont {Oudovenko}}, \bibinfo
  {author} {\bibfnamefont {O.}~\bibnamefont {Parcollet}}, \ and\ \bibinfo
  {author} {\bibfnamefont {C.~A.}\ \bibnamefont {Marianetti}},\ }\href
  {\doibase 10.1103/RevModPhys.78.865} {\bibfield  {journal} {\bibinfo
  {journal} {Review of Modern Physics}\ }
  \textbf {\bibinfo {volume} {78}},\ \bibinfo
  {pages} {865} (\bibinfo {year} {2006})}\BibitemShut {NoStop}%
\bibitem [{\citenamefont {Wills}\ \emph {et~al.}(2010)\citenamefont {Wills},
  \citenamefont {Alouani}, \citenamefont {Andersson}, \citenamefont {Delin},
  \citenamefont {Eriksson},\ and\ \citenamefont {Grechnev}}]{wills:2010th}%
  \BibitemOpen
  \bibfield  {author} {\bibinfo {author} {\bibfnamefont {J.~M.}\ \bibnamefont
  {Wills}}, \bibinfo {author} {\bibfnamefont {M.}~\bibnamefont {Alouani}},
  \bibinfo {author} {\bibfnamefont {P.}~\bibnamefont {Andersson}}, \bibinfo
  {author} {\bibfnamefont {A.}~\bibnamefont {Delin}}, \bibinfo {author}
  {\bibfnamefont {O.}~\bibnamefont {Eriksson}}, \ and\ \bibinfo {author}
  {\bibfnamefont {O.}~\bibnamefont {Grechnev}},\ }\href@noop {} {\bibfield
  {journal} {\bibinfo  {journal} {Springer Series in Solid-State Sciences}\ ,\
  \bibinfo {pages} {1}} (\bibinfo {year} {2010})}\BibitemShut {NoStop}%
\bibitem [{\citenamefont {Grechnev}\ \emph {et~al.}(2007)\citenamefont
  {Grechnev}, \citenamefont {Di~Marco}, \citenamefont {Katsnelson},
  \citenamefont {Lichtenstein}, \citenamefont {Wills},\ and\ \citenamefont
  {Eriksson}}]{Grechnev:2007en}%
  \BibitemOpen
  \bibfield  {author} {\bibinfo {author} {\bibfnamefont {A.}~\bibnamefont
  {Grechnev}}, \bibinfo {author} {\bibfnamefont {I.}~\bibnamefont {Di~Marco}},
  \bibinfo {author} {\bibfnamefont {M.~I.}\ \bibnamefont {Katsnelson}},
  \bibinfo {author} {\bibfnamefont {A.~I.}\ \bibnamefont {Lichtenstein}},
  \bibinfo {author} {\bibfnamefont {J.}~\bibnamefont {Wills}}, \ and\ \bibinfo
  {author} {\bibfnamefont {O.}~\bibnamefont {Eriksson}},\ }\href@noop {}
  {\bibfield  {journal} {\bibinfo  {journal} {Physical Review B}\ }\textbf
  {\bibinfo {volume} {76}},\ \bibinfo {pages} {035107} (\bibinfo {year}
  {2007})}\BibitemShut {NoStop}%
\bibitem [{\citenamefont {Gr{\aa}n{\"a}s}\ \emph {et~al.}(2012)\citenamefont
  {Gr{\aa}n{\"a}s}, \citenamefont {Di~Marco}, \citenamefont {Thunstr{\"o}m},
  \citenamefont {Nordstr{\"o}m}, \citenamefont {Eriksson}, \citenamefont
  {Bj{\"o}rkman},\ and\ \citenamefont {Wills}}]{Granas:2012hg}%
  \BibitemOpen
  \bibfield  {author} {\bibinfo {author} {\bibfnamefont {O.}~\bibnamefont
  {Gr{\aa}n{\"a}s}}, \bibinfo {author} {\bibfnamefont {I.}~\bibnamefont
  {Di~Marco}}, \bibinfo {author} {\bibfnamefont {P.}~\bibnamefont
  {Thunstr{\"o}m}}, \bibinfo {author} {\bibfnamefont {L.}~\bibnamefont
  {Nordstr{\"o}m}}, \bibinfo {author} {\bibfnamefont {O.}~\bibnamefont
  {Eriksson}}, \bibinfo {author} {\bibfnamefont {T.}~\bibnamefont
  {Bj{\"o}rkman}}, \ and\ \bibinfo {author} {\bibfnamefont {J.~M.}\
  \bibnamefont {Wills}},\ }\href@noop {} {\bibfield  {journal} {\bibinfo
  {journal} {Computational Materials Science}\ }\textbf {\bibinfo {volume}
  {55}},\ \bibinfo {pages} {295} (\bibinfo {year} {2012})}\BibitemShut
  {NoStop}%
\bibitem [{rsp()}]{rspt}%
  \BibitemOpen
  \href@noop {} {}\bibinfo {note}
  {\url{http://www.fplmto-rspt.org/}}\BibitemShut {NoStop}%
\bibitem [{\citenamefont {Wills}\ and\ \citenamefont
  {Cooper}(1987)}]{Wills:1987vt}%
  \BibitemOpen
  \bibfield  {author} {\bibinfo {author} {\bibfnamefont {J.~M.}\ \bibnamefont
  {Wills}}\ and\ \bibinfo {author} {\bibfnamefont {B.~R.}\ \bibnamefont
  {Cooper}},\ }\href {\doibase 10.1103/PhysRevB.36.3809} {\bibfield  {journal}
  {\bibinfo  {journal} {Physical Review B}\ }\textbf {\bibinfo {volume} {36}},\
  \bibinfo {pages} {3809} (\bibinfo {year} {1987})}\BibitemShut {NoStop}%
\bibitem [{\citenamefont {Blochl}\ \emph {et~al.}(1994)\citenamefont {Blochl},
  \citenamefont {Jepsen},\ and\ \citenamefont {Andersen}}]{Blochl:1994vg}%
  \BibitemOpen
  \bibfield  {author} {\bibinfo {author} {\bibfnamefont {P.~E.}\ \bibnamefont
  {Blochl}}, \bibinfo {author} {\bibfnamefont {O.}~\bibnamefont {Jepsen}}, \
  and\ \bibinfo {author} {\bibfnamefont {O.~K.}~\bibnamefont {Andersen}},\
  }\href@noop {} {\bibfield  {journal} {\bibinfo  {journal} {Physical Review
  B}\ }\textbf {\bibinfo {volume} {49}},\ \bibinfo {pages} {16223} (\bibinfo
  {year} {1994})}\BibitemShut {NoStop}%
\bibitem [{\citenamefont {Barth}\ and\ \citenamefont
  {Hedin}(1972)}]{Barth:1972un}%
  \BibitemOpen
  \bibfield  {author} {\bibinfo {author} {\bibfnamefont {U.}~\bibnamefont
  {Barth}}\ and\ \bibinfo {author} {\bibfnamefont {L.}~\bibnamefont {Hedin}},\
  }\href@noop {} {\bibfield  {journal} {\bibinfo  {journal} 
  {Journal of Physics C: Solid State Physics}\ }
  \textbf {\bibinfo {volume} {5}},\ \bibinfo {pages} {1629} (\bibinfo
  {year} {1972})}\BibitemShut {NoStop}%
\bibitem [{\citenamefont {Perdew}\ and\ \citenamefont
  {Wang}(1992)}]{Perdew:1992vq}%
  \BibitemOpen
  \bibfield  {author} {\bibinfo {author} {\bibfnamefont {J.~P.}~\bibnamefont
  {Perdew}}\ and\ \bibinfo {author} {\bibfnamefont {Y.}~\bibnamefont {Wang}},\
  }\href@noop {} {\bibfield  {journal} {\bibinfo  {journal} {Physical Review
  B}\ }\textbf {\bibinfo {volume} {45}},\ \bibinfo {pages} {13244}
  (\bibinfo {year} {1992})}\BibitemShut {NoStop}%
\bibitem [{\citenamefont {Perdew}\ and\ \citenamefont
  {Zunger}(1981)}]{Perdew:1981wm}%
  \BibitemOpen
  \bibfield  {author} {\bibinfo {author} {\bibfnamefont {J.~P.}~\bibnamefont
  {Perdew}}\ and\ \bibinfo {author} {\bibfnamefont {A.}~\bibnamefont
  {Zunger}},\ }\href@noop {} {\bibfield  {journal} {\bibinfo  {journal}
  {Physical Review B}\ }\textbf {\bibinfo {volume} {23}},\ \bibinfo {pages} {5048}
  (\bibinfo {year} {1981})}\BibitemShut {NoStop}%
\bibitem [{\citenamefont {Armiento}\ and\ \citenamefont
  {Mattsson}(2005)}]{Armiento:2005fk}%
  \BibitemOpen
  \bibfield  {author} {\bibinfo {author} {\bibfnamefont {R.}~\bibnamefont
  {Armiento}}\ and\ \bibinfo {author} {\bibfnamefont {A.~E.}\ \bibnamefont
  {Mattsson}},\ }\href {http://link.aps.org/doi/10.1103/PhysRevB.72.085108}
  {\bibfield  {journal} {\bibinfo  {journal} {Physical Review B}\ }\textbf
  {\bibinfo {volume} {72}},\ \bibinfo {pages} {085108} (\bibinfo {year}
  {2005})}\BibitemShut {NoStop}%
\bibitem [{\citenamefont {Perdew}\ \emph {et~al.}(1996)\citenamefont {Perdew},
  \citenamefont {Burke},\ and\ \citenamefont {Ernzerhof}}]{Perdew:1996uq}%
  \BibitemOpen
  \bibfield  {author} {\bibinfo {author} {\bibfnamefont {J.~P.}\ \bibnamefont
  {Perdew}}, \bibinfo {author} {\bibfnamefont {K.}~\bibnamefont {Burke}}, \
  and\ \bibinfo {author} {\bibfnamefont {M.}~\bibnamefont {Ernzerhof}},\ }\href
  {http://link.aps.org/doi/10.1103/PhysRevLett.77.3865} {\bibfield  {journal}
  {\bibinfo  {journal} {Physical Review Letters}\ }\textbf {\bibinfo {volume}
  {77}},\ \bibinfo {pages} {3865} (\bibinfo {year} {1996})}\BibitemShut
  {NoStop}%
\bibitem [{\citenamefont {Pourovskii}\ \emph {et~al.}(2005)\citenamefont
  {Pourovskii}, \citenamefont {Katsnelson},\ and\ \citenamefont
  {Lichtenstein}}]{Pourovskii:2005uw}%
  \BibitemOpen
  \bibfield  {author} {\bibinfo {author} {\bibfnamefont {L.~V.}~\bibnamefont
  {Pourovskii}}, \bibinfo {author} {\bibfnamefont {M.~I.}~\bibnamefont
  {Katsnelson}}, \ and\ \bibinfo {author} {\bibfnamefont {A.~I.}~\bibnamefont
  {Lichtenstein}},\ }\href@noop {} {\bibfield  {journal} {\bibinfo  {journal}
  {Physical Review B}\ }\textbf {\bibinfo {volume} {72}},\ \bibinfo {pages}
  {115106} (\bibinfo {year} {2005})}\BibitemShut {NoStop}%
\bibitem [{\citenamefont {Gr{\aa}n{\"a}s}(2012)}]{newFLEX}%
  \BibitemOpen
  \bibfield  {author} {\bibinfo {author} {\bibfnamefont {O.}~\bibnamefont
  {Gr{\aa}n{\"a}s}},\ }\emph {\bibinfo {title} {Theoretical Studies of
  Magnetism and Electron Correlation in Solids}},\ \href@noop {} {Ph.D.
  thesis},\ \bibinfo  {school} {Uppsala University, Materials Theory} (\bibinfo
  {year} {2012})\BibitemShut {NoStop}%
\bibitem [{\citenamefont {Bickers}\ and\ \citenamefont
  {Scalapino}(1989)}]{Bickers:1989fk}%
  \BibitemOpen
  \bibfield  {author} {\bibinfo {author} {\bibfnamefont {N.~E.}\ \bibnamefont
  {Bickers}}\ and\ \bibinfo {author} {\bibfnamefont {D.~J.}\ \bibnamefont
  {Scalapino}},\ }\href {\doibase
  http://dx.doi.org/10.1016/0003-4916(89)90359-X} {\bibfield  {journal}
  {\bibinfo  {journal} {Annals of Physics}\ }\textbf {\bibinfo {volume}
  {193}},\ \bibinfo {pages} {206} (\bibinfo {year} {1989})}\BibitemShut
  {NoStop}%
\bibitem [{Note1()}]{Note1}%
  \BibitemOpen
  \bibinfo {note} {Notice that the significance of the $U/W$ ratio to classify
  the strength of correlations is a bit diminished for multi-orbital systems
  where the corresponding orbitals are associated to different bandwidths and
  different matrix elements of the Coulomb interaction. In the main text we
  refer to this ratio only to to give an idea of the regime of applicability of
  our solver in terms of standard classifications.}\BibitemShut {Stop}%
\bibitem [{\citenamefont {Lichtenstein}\ \emph {et~al.}(2001)\citenamefont
  {Lichtenstein}, \citenamefont {Katsnelson},\ and\ \citenamefont
  {Kotliar}}]{Lichtenstein:2001uq}%
  \BibitemOpen
  \bibfield  {author} {\bibinfo {author} {\bibfnamefont {A.~I.}~\bibnamefont
  {Lichtenstein}}, \bibinfo {author} {\bibfnamefont {M.~I.}~\bibnamefont
  {Katsnelson}}, \ and\ \bibinfo {author} {\bibfnamefont {G.}~\bibnamefont
  {Kotliar}},\ }\href@noop {} {\bibfield  {journal} {\bibinfo  {journal} 
  {Physical Review Letters}\ }\textbf {\bibinfo {volume} {87}},\ \bibinfo {pages} {067205}
  (\bibinfo {year} {2001})}\BibitemShut {NoStop}%
\bibitem [{\citenamefont {{\c S}a{\c s}ıo{\u g}lu}\ \emph
  {et~al.}(2011)\citenamefont {{\c S}a{\c s}ıo{\u g}lu}, \citenamefont
  {Friedrich},\ and\ \citenamefont {Bl{\"u}gel}}]{Sasoglu:2011ch}%
  \BibitemOpen
  \bibfield  {author} {\bibinfo {author} {\bibfnamefont {E.}~\bibnamefont {{\c
  S}a{\c s}io{\u g}lu}}, \bibinfo {author} {\bibfnamefont {C.}~\bibnamefont
  {Friedrich}}, \ and\ \bibinfo {author} {\bibfnamefont {S.}~\bibnamefont
  {Bl{\"u}gel}},\ }\href@noop {} {\bibfield  {journal} {\bibinfo  {journal}
  {Physical Review B}\ }\textbf {\bibinfo {volume} {83}},\ \bibinfo {pages} {121101}
  (\bibinfo {year} {2011})}\BibitemShut {NoStop}%
\bibitem [{\citenamefont {Shai}\ \emph {et~al.}(2013)\citenamefont {Shai},
  \citenamefont {Adamo}, \citenamefont {Shen}, \citenamefont {Brooks},
  \citenamefont {Harter}, \citenamefont {Monkman}, \citenamefont {Burganov},
  \citenamefont {Schlom},\ and\ \citenamefont {Shen}}]{Shai2013}%
  \BibitemOpen
  \bibfield  {author} {\bibinfo {author} {\bibfnamefont {D.~E.}\ \bibnamefont
  {Shai}}, \bibinfo {author} {\bibfnamefont {C.}~\bibnamefont {Adamo}},
  \bibinfo {author} {\bibfnamefont {D.~W.}\ \bibnamefont {Shen}}, \bibinfo
  {author} {\bibfnamefont {C.~M.}\ \bibnamefont {Brooks}}, \bibinfo {author}
  {\bibfnamefont {J.~W.}\ \bibnamefont {Harter}}, \bibinfo {author}
  {\bibfnamefont {E.~J.}\ \bibnamefont {Monkman}}, \bibinfo {author}
  {\bibfnamefont {B.}~\bibnamefont {Burganov}}, \bibinfo {author}
  {\bibfnamefont {D.~G.}\ \bibnamefont {Schlom}}, \ and\ \bibinfo {author}
  {\bibfnamefont {K.~M.}\ \bibnamefont {Shen}},\ }\href {\doibase
  10.1103/PhysRevLett.110.087004} {\bibfield  {journal} {\bibinfo  {journal}
  {Physical Review Letters}\ }\textbf {\bibinfo {volume} {110}},\ \bibinfo {pages}
  {087004} (\bibinfo {year} {2013})}\BibitemShut {NoStop}%
\bibitem [{\citenamefont {Bushmeleva}\ \emph {et~al.}(2006)\citenamefont
  {Bushmeleva}, \citenamefont {Pomjakushin}, \citenamefont {Pomjakushina},
  \citenamefont {Sheptyakov},\ and\ \citenamefont
  {Balagurov}}]{Bushmeleva:2006jn}%
  \BibitemOpen
  \bibfield  {author} {\bibinfo {author} {\bibfnamefont {S.~N.}\ \bibnamefont
  {Bushmeleva}}, \bibinfo {author} {\bibfnamefont {V.~Y.}\ \bibnamefont
  {Pomjakushin}}, \bibinfo {author} {\bibfnamefont {E.~V.}\ \bibnamefont
  {Pomjakushina}}, \bibinfo {author} {\bibfnamefont {D.~V.}\ \bibnamefont
  {Sheptyakov}}, \ and\ \bibinfo {author} {\bibfnamefont {A.~M.}\ \bibnamefont
  {Balagurov}},\ }\href@noop {} {\bibfield  {journal} {\bibinfo  {journal} 
  {Journal of Magnetism and Magnetic Materials}\ }
  \textbf {\bibinfo {volume} {305}},\ \bibinfo {pages} {491}
  (\bibinfo {year} {2006})}\BibitemShut {NoStop}%
\bibitem [{\citenamefont {Maiti}\ and\ \citenamefont
  {Singh}(2005)}]{Maiti2005}%
  \BibitemOpen
  \bibfield  {author} {\bibinfo {author} {\bibfnamefont {K.}~\bibnamefont
  {Maiti}}\ and\ \bibinfo {author} {\bibfnamefont {R.~S.}\ \bibnamefont
  {Singh}},\ }\href {http://link.aps.org/doi/10.1103/PhysRevB.71.161102}
  {\bibfield  {journal} {\bibinfo  {journal} {Physcal Review B}\ }\textbf {\bibinfo
  {volume} {71}},\ \bibinfo {pages} {161102} (\bibinfo {year}
  {2005})}\BibitemShut {NoStop}%
\bibitem [{\citenamefont {Lee}\ \emph {et~al.}(2001)\citenamefont {Lee},
  \citenamefont {Lee}, \citenamefont {Noh}, \citenamefont {Char}, \citenamefont
  {Park}, \citenamefont {Oh}, \citenamefont {Park}, \citenamefont {Eom},
  \citenamefont {Takeda},\ and\ \citenamefont {Kanno}}]{Lee:2001kx}%
  \BibitemOpen
  \bibfield  {author} {\bibinfo {author} {\bibfnamefont {J.~S.}\ \bibnamefont
  {Lee}}, \bibinfo {author} {\bibfnamefont {Y.~S.}\ \bibnamefont {Lee}},
  \bibinfo {author} {\bibfnamefont {T.~W.}\ \bibnamefont {Noh}}, \bibinfo
  {author} {\bibfnamefont {K.}~\bibnamefont {Char}}, \bibinfo {author}
  {\bibfnamefont {J.}~\bibnamefont {Park}}, \bibinfo {author} {\bibfnamefont
  {S.~J.}\ \bibnamefont {Oh}}, \bibinfo {author} {\bibfnamefont {J.~H.}\
  \bibnamefont {Park}}, \bibinfo {author} {\bibfnamefont {C.~B.}\ \bibnamefont
  {Eom}}, \bibinfo {author} {\bibfnamefont {T.}~\bibnamefont {Takeda}}, \ and\
  \bibinfo {author} {\bibfnamefont {R.}~\bibnamefont {Kanno}},\ }\href
  {http://link.aps.org/doi/10.1103/PhysRevB.64.245107} {\bibfield  {journal}
  {\bibinfo  {journal} {Physical Review B}\ }\textbf {\bibinfo {volume} {64}},\
  \bibinfo {pages} {245107} (\bibinfo {year} {2001})}\BibitemShut {NoStop}%
\bibitem [{\citenamefont {Etz}\ \emph {et~al.}(2012)\citenamefont {Etz},
  \citenamefont {Maznichenko}, \citenamefont {B{\"o}ttcher}, \citenamefont
  {Henk}, \citenamefont {Yaresko}, \citenamefont {Hergert}, \citenamefont
  {Mazin}, \citenamefont {Mertig},\ and\ \citenamefont {Ernst}}]{Etz:2012uq}%
  \BibitemOpen
  \bibfield  {author} {\bibinfo {author} {\bibfnamefont {C.}~\bibnamefont
  {Etz}}, \bibinfo {author} {\bibfnamefont {I.~V.}\ \bibnamefont
  {Maznichenko}}, \bibinfo {author} {\bibfnamefont {D.}~\bibnamefont
  {B{\"o}ttcher}}, \bibinfo {author} {\bibfnamefont {J.}~\bibnamefont {Henk}},
  \bibinfo {author} {\bibfnamefont {A.~N.}\ \bibnamefont {Yaresko}}, \bibinfo
  {author} {\bibfnamefont {W.}~\bibnamefont {Hergert}}, \bibinfo {author}
  {\bibfnamefont {I.~I.}\ \bibnamefont {Mazin}}, \bibinfo {author}
  {\bibfnamefont {I.}~\bibnamefont {Mertig}}, \ and\ \bibinfo {author}
  {\bibfnamefont {A.}~\bibnamefont {Ernst}},\ }\href
  {http://link.aps.org/doi/10.1103/PhysRevB.86.064441} {\bibfield  {journal}
  {\bibinfo  {journal} {Physical Review B}\ }\textbf {\bibinfo {volume} {86}},\
  \bibinfo {pages} {064441} (\bibinfo {year} {2012})}\BibitemShut {NoStop}%
\bibitem [{\citenamefont {Zayak}\ \emph {et~al.}(2006)\citenamefont {Zayak},
  \citenamefont {Huang}, \citenamefont {Neaton},\ and\ \citenamefont
  {Rabe}}]{Zayak2006}%
  \BibitemOpen
  \bibfield  {author} {\bibinfo {author} {\bibfnamefont {A.~T.}\ \bibnamefont
  {Zayak}}, \bibinfo {author} {\bibfnamefont {X.}~\bibnamefont {Huang}},
  \bibinfo {author} {\bibfnamefont {J.~B.}\ \bibnamefont {Neaton}}, \ and\
  \bibinfo {author} {\bibfnamefont {K.~M.}\ \bibnamefont {Rabe}},\ }\href
  {http://link.aps.org/doi/10.1103/PhysRevB.74.094104} {\bibfield  {journal}
  {\bibinfo  {journal} {Physical Review B}\ }\textbf {\bibinfo {volume} {74}},\
  \bibinfo {pages} {094104} (\bibinfo {year} {2006})}\BibitemShut {NoStop}%
\bibitem [{\citenamefont {Vinet}\ \emph {et~al.}(1989)\citenamefont {Vinet},
  \citenamefont {Rose}, \citenamefont {Ferrante},\ and\ \citenamefont
  {Smith}}]{Vinet:1989fv}%
  \BibitemOpen
  \bibfield  {author} {\bibinfo {author} {\bibfnamefont {P.}~\bibnamefont
  {Vinet}}, \bibinfo {author} {\bibfnamefont {J.~H.}\ \bibnamefont {Rose}},
  \bibinfo {author} {\bibfnamefont {J.}~\bibnamefont {Ferrante}}, \ and\
  \bibinfo {author} {\bibfnamefont {J.~R.}\ \bibnamefont {Smith}},\ }\href
  {http://stacks.iop.org/0953-8984/1/i=11/a=002} {\bibfield  {journal}
  {\bibinfo  {journal} {Journal of Physics: Condensed Matter}\ }\textbf
  {\bibinfo {volume} {1}},\ \bibinfo {pages} {1941} (\bibinfo {year}
  {1989})}\BibitemShut {NoStop}%
\bibitem [{\citenamefont {Hamlin}\ \emph {et~al.}(2007)\citenamefont {Hamlin},
  \citenamefont {Deemyad}, \citenamefont {Schilling}, \citenamefont {Jacobsen},
  \citenamefont {Kumar}, \citenamefont {Cornelius}, \citenamefont {Cao},\ and\
  \citenamefont {Neumeier}}]{Hamlin:2007ct}%
  \BibitemOpen
  \bibfield  {author} {\bibinfo {author} {\bibfnamefont {J.~J.}\ \bibnamefont
  {Hamlin}}, \bibinfo {author} {\bibfnamefont {S.}~\bibnamefont {Deemyad}},
  \bibinfo {author} {\bibfnamefont {J.~S.}\ \bibnamefont {Schilling}}, \bibinfo
  {author} {\bibfnamefont {M.~K.}\ \bibnamefont {Jacobsen}}, \bibinfo {author}
  {\bibfnamefont {R.~S.}\ \bibnamefont {Kumar}}, \bibinfo {author}
  {\bibfnamefont {A.~L.}\ \bibnamefont {Cornelius}}, \bibinfo {author}
  {\bibfnamefont {G.}~\bibnamefont {Cao}}, \ and\ \bibinfo {author}
  {\bibfnamefont {J.~J.}\ \bibnamefont {Neumeier}},\ }\href
  {http://link.aps.org/doi/10.1103/PhysRevB.76.014432} {\bibfield  {journal}
  {\bibinfo  {journal} {Physical Review B}\ }\textbf {\bibinfo {volume} {76}},\
  \bibinfo {pages} {014432} (\bibinfo {year} {2007})}\BibitemShut {NoStop}%
\bibitem [{\citenamefont {Gu}\ \emph {et~al.}(2012)\citenamefont {Gu},
  \citenamefont {Xie}, \citenamefont {Shen}, \citenamefont {Xie}, \citenamefont
  {Wang}, \citenamefont {Tang}, \citenamefont {Wu}, \citenamefont {Zhang},\
  and\ \citenamefont {Wu}}]{Gu:2012uq}%
  \BibitemOpen
  \bibfield  {author} {\bibinfo {author} {\bibfnamefont {M.}~\bibnamefont
  {Gu}}, \bibinfo {author} {\bibfnamefont {Q.}~\bibnamefont {Xie}}, \bibinfo
  {author} {\bibfnamefont {X.}~\bibnamefont {Shen}}, \bibinfo {author}
  {\bibfnamefont {R.}~\bibnamefont {Xie}}, \bibinfo {author} {\bibfnamefont
  {J.}~\bibnamefont {Wang}}, \bibinfo {author} {\bibfnamefont {G.}~\bibnamefont
  {Tang}}, \bibinfo {author} {\bibfnamefont {D.}~\bibnamefont {Wu}}, \bibinfo
  {author} {\bibfnamefont {G.~P.}\ \bibnamefont {Zhang}}, \ and\ \bibinfo
  {author} {\bibfnamefont {X.~S.}\ \bibnamefont {Wu}},\ }\href
  {http://link.aps.org/doi/10.1103/PhysRevLett.109.157003} {\bibfield
  {journal} {\bibinfo  {journal} {Physical Review Letters}\ }\textbf {\bibinfo
  {volume} {109}},\ \bibinfo {pages} {157003} (\bibinfo {year}
  {2012})}\BibitemShut {NoStop}%
\bibitem [{\citenamefont {Takizawa}\ \emph {et~al.}(2005)\citenamefont
  {Takizawa}, \citenamefont {Toyota}, \citenamefont {Wadati}, \citenamefont
  {Chikamatsu}, \citenamefont {Kumigashira}, \citenamefont {Fujimori},
  \citenamefont {Oshima}, \citenamefont {Fang}, \citenamefont {Lippmaa},
  \citenamefont {Kawasaki},\ and\ \citenamefont {Koinuma}}]{Takizawa:2005vn}%
  \BibitemOpen
  \bibfield  {author} {\bibinfo {author} {\bibfnamefont {M.}~\bibnamefont
  {Takizawa}}, \bibinfo {author} {\bibfnamefont {D.}~\bibnamefont {Toyota}},
  \bibinfo {author} {\bibfnamefont {H.}~\bibnamefont {Wadati}}, \bibinfo
  {author} {\bibfnamefont {A.}~\bibnamefont {Chikamatsu}}, \bibinfo {author}
  {\bibfnamefont {H.}~\bibnamefont {Kumigashira}}, \bibinfo {author}
  {\bibfnamefont {A.}~\bibnamefont {Fujimori}}, \bibinfo {author}
  {\bibfnamefont {M.}~\bibnamefont {Oshima}}, \bibinfo {author} {\bibfnamefont
  {Z.}~\bibnamefont {Fang}}, \bibinfo {author} {\bibfnamefont {M.}~\bibnamefont
  {Lippmaa}}, \bibinfo {author} {\bibfnamefont {M.}~\bibnamefont {Kawasaki}}, \
  and\ \bibinfo {author} {\bibfnamefont {H.}~\bibnamefont {Koinuma}},\ }\href
  {http://link.aps.org/doi/10.1103/PhysRevB.72.060404} {\bibfield  {journal}
  {\bibinfo  {journal} {Physical Review B}\ }\textbf {\bibinfo {volume} {72}},\
  \bibinfo {pages} {060404} (\bibinfo {year} {2005})}\BibitemShut {NoStop}%
\bibitem [{\citenamefont {Toyota}\ \emph {et~al.}(2005)\citenamefont {Toyota},
  \citenamefont {Ohkubo}, \citenamefont {Kumigashira}, \citenamefont {Oshima},
  \citenamefont {Ohnishi}, \citenamefont {Lippmaa}, \citenamefont {Takizawa},
  \citenamefont {Fujimori}, \citenamefont {Ono}, \citenamefont {Kawasaki},\
  and\ \citenamefont {Koinuma}}]{Toyota:2005ge}%
  \BibitemOpen
  \bibfield  {author} {\bibinfo {author} {\bibfnamefont {D.}~\bibnamefont
  {Toyota}}, \bibinfo {author} {\bibfnamefont {I.}~\bibnamefont {Ohkubo}},
  \bibinfo {author} {\bibfnamefont {H.}~\bibnamefont {Kumigashira}}, \bibinfo
  {author} {\bibfnamefont {M.}~\bibnamefont {Oshima}}, \bibinfo {author}
  {\bibfnamefont {T.}~\bibnamefont {Ohnishi}}, \bibinfo {author} {\bibfnamefont
  {M.}~\bibnamefont {Lippmaa}}, \bibinfo {author} {\bibfnamefont
  {M.}~\bibnamefont {Takizawa}}, \bibinfo {author} {\bibfnamefont
  {A.}~\bibnamefont {Fujimori}}, \bibinfo {author} {\bibfnamefont
  {K.}~\bibnamefont {Ono}}, \bibinfo {author} {\bibfnamefont {M.}~\bibnamefont
  {Kawasaki}}, \ and\ \bibinfo {author} {\bibfnamefont {H.}~\bibnamefont
  {Koinuma}},\ }\href@noop {} {\bibfield  {journal} {\bibinfo  {journal} 
  {Applied Physics Letters}\ }
  \textbf {\bibinfo {volume} {87}},\ \bibinfo {pages} {162508}
  (\bibinfo {year} {2005})}\BibitemShut {NoStop}%
\bibitem [{\citenamefont {Alexander}\ \emph {et~al.}(2005)\citenamefont
  {Alexander}, \citenamefont {McCall}, \citenamefont {Schlottmann},
  \citenamefont {Crow},\ and\ \citenamefont {Cao}}]{Alexander:2005iu}%
  \BibitemOpen
  \bibfield  {author} {\bibinfo {author} {\bibfnamefont {C.~S.}~\bibnamefont
  {Alexander}}, \bibinfo {author} {\bibfnamefont {S.}~\bibnamefont {McCall}},
  \bibinfo {author} {\bibfnamefont {P.}~\bibnamefont {Schlottmann}}, \bibinfo
  {author} {\bibfnamefont {J.~E.}~\bibnamefont {Crow}}, \ and\ \bibinfo {author}
  {\bibfnamefont {G.}~\bibnamefont {Cao}},\ }\href@noop {} {\bibfield
  {journal} {\bibinfo  {journal} {Physical Review B}\ }\textbf {\bibinfo
  {volume} {72}},\ \bibinfo {pages} {024415}
  (\bibinfo {year} {2005})}\BibitemShut {NoStop}%
\bibitem [{\citenamefont {Allen}\ \emph {et~al.}(1996)\citenamefont {Allen},
  \citenamefont {Berger}, \citenamefont {Chauvet}, \citenamefont {Forro},
  \citenamefont {Jarlborg}, \citenamefont {Junod}, \citenamefont {Revaz},\ and\
  \citenamefont {Santi}}]{Allen1996}%
  \BibitemOpen
  \bibfield  {author} {\bibinfo {author} {\bibfnamefont {P.~B.}\ \bibnamefont
  {Allen}}, \bibinfo {author} {\bibfnamefont {H.}~\bibnamefont {Berger}},
  \bibinfo {author} {\bibfnamefont {O.}~\bibnamefont {Chauvet}}, \bibinfo
  {author} {\bibfnamefont {L.}~\bibnamefont {Forro}}, \bibinfo {author}
  {\bibfnamefont {T.}~\bibnamefont {Jarlborg}}, \bibinfo {author}
  {\bibfnamefont {A.}~\bibnamefont {Junod}}, \bibinfo {author} {\bibfnamefont
  {B.}~\bibnamefont {Revaz}}, \ and\ \bibinfo {author} {\bibfnamefont
  {G.}~\bibnamefont {Santi}},\ }\href
  {http://link.aps.org/doi/10.1103/PhysRevB.53.4393} {\bibfield  {journal}
  {\bibinfo  {journal} {Physical Review B}\ }\textbf {\bibinfo {volume} {53}},\
  \bibinfo {pages} {4393} (\bibinfo {year} {1996})}\BibitemShut {NoStop}%
\bibitem [{\citenamefont {Okamoto}\ \emph {et~al.}(1999)\citenamefont
  {Okamoto}, \citenamefont {Mizokawa}, \citenamefont {Fujimori}, \citenamefont
  {Hase}, \citenamefont {Nohara}, \citenamefont {Takagi}, \citenamefont
  {Takeda},\ and\ \citenamefont {Takano}}]{Okamoto:1999wk}%
  \BibitemOpen
  \bibfield  {author} {\bibinfo {author} {\bibfnamefont {J.}~\bibnamefont
  {Okamoto}}, \bibinfo {author} {\bibfnamefont {T.}~\bibnamefont {Mizokawa}},
  \bibinfo {author} {\bibfnamefont {A.}~\bibnamefont {Fujimori}}, \bibinfo
  {author} {\bibfnamefont {I.}~\bibnamefont {Hase}}, \bibinfo {author}
  {\bibfnamefont {M.}~\bibnamefont {Nohara}}, \bibinfo {author} {\bibfnamefont
  {H.}~\bibnamefont {Takagi}}, \bibinfo {author} {\bibfnamefont
  {Y.}~\bibnamefont {Takeda}}, \ and\ \bibinfo {author} {\bibfnamefont
  {M.}~\bibnamefont {Takano}},\ }\href@noop {} {\bibfield  {journal} {\bibinfo
  {journal} {Physical Review B}\ }\textbf {\bibinfo {volume} {60}},\ \bibinfo
  {pages} {2281} (\bibinfo {year} {1999})}\BibitemShut {NoStop}%
\bibitem [{\citenamefont {Nekrasov}\ \emph {et~al.}(2006)\citenamefont {Nekrasov},
  \citenamefont {Held}, \citenamefont {Keller},\citenamefont {Kondakov},
  \citenamefont {Pruschke},
  \citenamefont {Kollar},
  \citenamefont {Andersen},
  \citenamefont {Anisimov},
  \ and\ \citenamefont {Vollhardt}}]{Nekrasov:2006ha}%
  \BibitemOpen
  \bibfield  {author} {\bibinfo {author} {\bibfnamefont {I.~A.}\ \bibnamefont
  {Nekrasov}}, \bibinfo {author} {\bibfnamefont {K.}\ \bibnamefont {Held}},
  \bibinfo {author} {\bibfnamefont {G.}~\bibnamefont {Keller}},
  \bibinfo {author} {\bibfnamefont {D.~E.}~\bibnamefont {Kondakov}},
  \bibinfo {author} {\bibfnamefont {Th.}~\bibnamefont {Pruschke}},
  \bibinfo {author} {\bibfnamefont {M.}~\bibnamefont {Kollar}},
  \bibinfo {author} {\bibfnamefont {O.~K.}~\bibnamefont {Andersen}},
  \bibinfo {author} {\bibfnamefont {V.~I.}~\bibnamefont {Anisimov}},
  \ and\ \bibinfo {author} {\bibfnamefont {D.}\ \bibnamefont {Vollhardt}},\ } 
  {http://link.aps.org/doi/10.1103/PhysRevB.73.155112}
  {\bibfield  {journal} {\bibinfo  {journal} {Physical Review B}\ }\textbf
  {\bibinfo {volume} {73}},\ \bibinfo {pages} {155122} 
  (\bibinfo {year} {2006})}\BibitemShut {NoStop}%
\bibitem [{\citenamefont {Held}\ \emph {et~al.}(2012)\citenamefont {Held},
  \citenamefont {Peters},\ and\ \citenamefont
  {Toschi}}]{Held:2013dq}%
  \BibitemOpen
  \bibfield  {author} {\bibinfo {author} {\bibfnamefont {K.}~\bibnamefont
  {Held}}, \bibinfo {author} {\bibfnamefont {R.}~\bibnamefont {Peters}}, \
  and\ \bibinfo {author} {\bibfnamefont {A.}~\bibnamefont {Toschi}},\ }\href@noop {}
  {\bibfield  {journal} {\bibinfo
  {journal} {Physical Review Letters}\ }\textbf {\bibinfo {volume} {110}},\ \bibinfo
  {pages} {246402} (\bibinfo {year} {2013})}\BibitemShut {NoStop}%
\bibitem [{\citenamefont {Liu}\ \emph {et~al.}(2008)\citenamefont {Liu},
  \citenamefont {Antonov}, \citenamefont {Jepsen},\ and\ \citenamefont
  {Andersen}}]{Liu:2008io}%
  \BibitemOpen
  \bibfield  {author} {\bibinfo {author} {\bibfnamefont {G.-Q.}\ \bibnamefont
  {Liu}}, \bibinfo {author} {\bibfnamefont {V.~N.}\ \bibnamefont {Antonov}},
  \bibinfo {author} {\bibfnamefont {O.}~\bibnamefont {Jepsen}}, \ and\ \bibinfo
  {author} {\bibfnamefont {O.~K.}\ \bibnamefont {Andersen}},\ }\href@noop {}
  {\bibfield  {journal} {\bibinfo  {journal} {Physcal Review Letters}\ }\textbf
  {\bibinfo {volume} {101}},\ \bibinfo {pages} {026408} 
  (\bibinfo {year} {2008})}\BibitemShut {NoStop}%
\bibitem [{\citenamefont {Yeh}\ and\ \citenamefont
  {Lindau}(1985)}]{Yeh:1985vt}%
  \BibitemOpen
  \bibfield  {author} {\bibinfo {author} {\bibfnamefont {J.}~\bibnamefont
  {Yeh}}\ and\ \bibinfo {author} {\bibfnamefont {I.}~\bibnamefont {Lindau}},\
  }\href {\doibase http://dx.doi.org/10.1016/0092-640X(85)90016-6} {\bibfield
  {journal} {\bibinfo  {journal} {Atomic Data and Nuclear Data Tables}\
  }\textbf {\bibinfo {volume} {32}},\ \bibinfo {pages} {1 } (\bibinfo {year}
  {1985})}\BibitemShut {NoStop}%
\bibitem [{\citenamefont {Casula}\ \emph {et~al.}(2012)\citenamefont {Casula},
  \citenamefont {Rubtsov},\ and\ \citenamefont
  {Biermann}}]{PhysRevB.85.035115}%
  \BibitemOpen
  \bibfield  {author} {\bibinfo {author} {\bibfnamefont {M.}~\bibnamefont
  {Casula}}, \bibinfo {author} {\bibfnamefont {A.}~\bibnamefont {Rubtsov}}, \
  and\ \bibinfo {author} {\bibfnamefont {S.}~\bibnamefont {Biermann}},\ }\href
  {\doibase 10.1103/PhysRevB.85.035115} {\bibfield  {journal} {\bibinfo
  {journal} {Physical Review B}\ }\textbf {\bibinfo {volume} {85}},\ \bibinfo
  {pages} {035115} (\bibinfo {year} {2012})}\BibitemShut {NoStop}%
\bibitem [{\citenamefont {Grutter}\ \emph {et~al.}(2012)\citenamefont
  {Grutter}, \citenamefont {Wong}, \citenamefont {Arenholz}, \citenamefont
  {Vailionis},\ and\ \citenamefont {Suzuki}}]{Grutter:2012bp}%
  \BibitemOpen
  \bibfield  {author} {\bibinfo {author} {\bibfnamefont {A.~J.}\ \bibnamefont
  {Grutter}}, \bibinfo {author} {\bibfnamefont {F.~J.}\ \bibnamefont {Wong}},
  \bibinfo {author} {\bibfnamefont {E.}~\bibnamefont {Arenholz}}, \bibinfo
  {author} {\bibfnamefont {A.}~\bibnamefont {Vailionis}}, \ and\ \bibinfo
  {author} {\bibfnamefont {Y.}~\bibnamefont {Suzuki}},\ }\href {\doibase
  10.1103/PhysRevB.85.134429} {\bibfield  {journal} {\bibinfo  {journal} 
  {Physical Review B}\ }
  \textbf {\bibinfo {volume} {85}},\ \bibinfo {pages} {134429}
  (\bibinfo {year} {2012})}\BibitemShut {NoStop}%
\bibitem [{\citenamefont {Solovyev}(1998)}]{Solovyev:1998ec}%
  \BibitemOpen
  \bibfield  {author} {\bibinfo {author} {\bibfnamefont {I.~V.}\ \bibnamefont
  {Solovyev}},\ }\href@noop {} {\bibfield  {journal} {\bibinfo  {journal} 
  {Journal of Magnetism and Magnetic Matererials}\ }
  \textbf {\bibinfo {volume} {177-181}},\ \bibinfo {pages}
  {811} (\bibinfo {year} {1998})}\BibitemShut {NoStop}%
\bibitem [{\citenamefont {Thole}\ \emph {et~al.}(1992)\citenamefont {Thole},
  \citenamefont {Carra}, \citenamefont {Sette},\ and\ \citenamefont {van~der
  Laan}}]{Thole:1992tm}%
  \BibitemOpen
  \bibfield  {author} {\bibinfo {author} {\bibfnamefont {B.~T.}\ \bibnamefont
  {Thole}}, \bibinfo {author} {\bibfnamefont {P.}~\bibnamefont {Carra}},
  \bibinfo {author} {\bibfnamefont {F.}~\bibnamefont {Sette}}, \ and\ \bibinfo
  {author} {\bibfnamefont {G.}~\bibnamefont {van~der Laan}},\ }\href
  {http://link.aps.org/doi/10.1103/PhysRevLett.68.1943} {\bibfield  {journal}
  {\bibinfo  {journal} {Physical Review Letters}\ }\textbf {\bibinfo {volume}
  {68}},\ \bibinfo {pages} {1943} (\bibinfo {year} {1992})}\BibitemShut
  {NoStop}%
\bibitem [{\citenamefont {Rusz}\ \emph {et~al.}(2011)\citenamefont {Rusz},
  \citenamefont {Rubino}, \citenamefont {Eriksson}, \citenamefont {Oppeneer},\
  and\ \citenamefont {Leifer}}]{Rusz:2011cf}%
  \BibitemOpen
  \bibfield  {author} {\bibinfo {author} {\bibfnamefont {J.}~\bibnamefont
  {Rusz}}, \bibinfo {author} {\bibfnamefont {S.}~\bibnamefont {Rubino}},
  \bibinfo {author} {\bibfnamefont {O.}~\bibnamefont {Eriksson}}, \bibinfo
  {author} {\bibfnamefont {P.~M.}\ \bibnamefont {Oppeneer}}, \ and\ \bibinfo
  {author} {\bibfnamefont {K.}~\bibnamefont {Leifer}},\ }\href@noop {}
  {\bibfield  {journal} {\bibinfo  {journal} {Physical Review B}\ }\textbf {\bibinfo
  {volume} {84}},\ \bibinfo {pages} {064444} (\bibinfo {year}
  {2011})}\BibitemShut {NoStop}%
\bibitem [{\citenamefont {Verbeeck}\ \emph {et~al.}(2010)\citenamefont
  {Verbeeck}, \citenamefont {Tian},\ and\ \citenamefont
  {Schattschneider}}]{Verbeeck:2010is}%
  \BibitemOpen
  \bibfield  {author} {\bibinfo {author} {\bibfnamefont {J.}~\bibnamefont
  {Verbeeck}}, \bibinfo {author} {\bibfnamefont {H.}~\bibnamefont {Tian}}, \
  and\ \bibinfo {author} {\bibfnamefont {P.}~\bibnamefont {Schattschneider}},\
  }\href@noop {} {\bibfield  {journal} {\bibinfo  {journal} {Nature}\ }\textbf
  {\bibinfo {volume} {467}},\ \bibinfo {pages} {301} (\bibinfo {year}
  {2010})}\BibitemShut {NoStop}%
\bibitem [{\citenamefont {Treves}(1962)}]{Treves:1962cs}%
  \BibitemOpen
  \bibfield  {author} {\bibinfo {author} {\bibfnamefont {D.}~\bibnamefont
  {Treves}},\ }\href {http://link.aps.org/doi/10.1103/PhysRev.125.1843}
  {\bibfield  {journal} {\bibinfo  {journal} {Physical Review}\ }\textbf
  {\bibinfo {volume} {125}},\ \bibinfo {pages} {1843} (\bibinfo {year}
  {1962})}\BibitemShut {NoStop}%
\bibitem [{\citenamefont {Solovyev}(1997)}]{Solovyev:1997bt}%
  \BibitemOpen
  \bibfield  {author} {\bibinfo {author} {\bibfnamefont {I.~V.}\ \bibnamefont
  {Solovyev}},\ }\href {http://link.aps.org/doi/10.1103/PhysRevB.55.8060}
  {\bibfield  {journal} {\bibinfo  {journal} {Physical Review B}\ }\textbf
  {\bibinfo {volume} {55}},\ \bibinfo {pages} {8060} (\bibinfo {year}
  {1997})}\BibitemShut {NoStop}%
\bibitem [{\citenamefont {D{\"u}rr}\ and\ \citenamefont {van~der
  Laan}(1996)}]{Durr:1996he}%
  \BibitemOpen
  \bibfield  {author} {\bibinfo {author} {\bibfnamefont {H.~A.}\ \bibnamefont
  {D{\"u}rr}}\ and\ \bibinfo {author} {\bibfnamefont {G.}~\bibnamefont {van~der
  Laan}},\ }\href {http://link.aps.org/doi/10.1103/PhysRevB.54.R760} {\bibfield
   {journal} {\bibinfo  {journal} {Physical Review B}\ }\textbf {\bibinfo
  {volume} {54}},\ \bibinfo {pages} {R760} (\bibinfo {year}
  {1996})}\BibitemShut {NoStop}%
\bibitem [{elk()}]{elk-calcs}%
  \BibitemOpen
  \href@noop {} {}\bibinfo {note} {A fully non-collinear calculation was done
  with the Elk code, \url{http://elk.sourceforge.net}. Showing a tilting of 
  3$^{\circ}$ for the experimental volume with the LDA functional by von Barth and
  Hedin. The reason the tilting is not larger is that the Ru d-shell is close
  to half filling, hence the coupling between spin and orbital moment is
  considerably smaller than the exchange coupling.}\BibitemShut {Stop}%
\bibitem [{\citenamefont {Bultmark}\ \emph {et~al.}(2009)\citenamefont
  {Bultmark}, \citenamefont {Cricchio}, \citenamefont {Gr\aa{}n\"as},\ and\
  \citenamefont {Nordstr\"om}}]{Bultmark:2009gm}%
  \BibitemOpen
  \bibfield  {author} {\bibinfo {author} {\bibfnamefont {F.}~\bibnamefont
  {Bultmark}}, \bibinfo {author} {\bibfnamefont {F.}~\bibnamefont {Cricchio}},
  \bibinfo {author} {\bibfnamefont {O.}~\bibnamefont {Gr\aa{}n\"as}}, \ and\
  \bibinfo {author} {\bibfnamefont {L.}~\bibnamefont {Nordstr\"om}},\ }\href
  {\doibase 10.1103/PhysRevB.80.035121} {\bibfield  {journal} {\bibinfo
  {journal} {Physical Review B}\ }\textbf {\bibinfo {volume} {80}},\ \bibinfo
  {pages} {035121} (\bibinfo {year} {2009})}\BibitemShut {NoStop}%
\bibitem [{\citenamefont {Bruno}(1989)}]{Bruno:1989th}%
  \BibitemOpen
  \bibfield  {author} {\bibinfo {author} {\bibfnamefont {P.}~\bibnamefont
  {Bruno}},\ }\href@noop {} {\bibfield  {journal} {\bibinfo  {journal} 
  {Physical Review B}\ }
  \textbf {\bibinfo {volume} {39}},\ \bibinfo {pages} {865} (\bibinfo
  {year} {1989})}\BibitemShut {NoStop}%
\end{thebibliography}%
\end{document}